\documentclass[letterpaper,twocolumn,pre,aps,showpacs,floatfix,superscriptaddress,nofootinbib]{revtex4-1}
\usepackage[latin1]{inputenc}
\usepackage{bm}
\usepackage[usenames,dvipsnames]{color}
\usepackage{multirow}
\usepackage{amssymb}
\usepackage{amsbsy}
\usepackage{amsmath}
\usepackage{stmaryrd}
\usepackage{graphicx}
\usepackage{epsfig}
\usepackage{placeins}
\usepackage{ulem}
\usepackage{soul}
\usepackage{braket}
\usepackage{blindtext}
\usepackage[colorlinks,linkcolor=blue,citecolor=blue,urlcolor=blue]{hyperref}
\makeatletter

\begin{document}

\title{Real-Time Dynamics of Typical and Untypical States in 
Non-Integrable Systems}

\author{Jonas Richter}\email{jonasrichter@uos.de} \affiliation{Departement of
Physics, University of Osnabr\"uck, D-49069 Osnabr\"uck,
Germany}

\author{Fengping Jin}
\affiliation{Institute for Advanced Simulation, J\"ulich Supercomputing Centre,  
Forschungszentrum J\"ulich, D-52425 J\"ulich, Germany}

\author{Hans De Raedt}
\affiliation{Zernike Institute for Advanced Materials, University of Groningen,  
NL-9747AG Groningen, The Netherlands}

\author{Kristel Michielsen}
\affiliation{Institute for Advanced Simulation, J\"ulich Supercomputing Centre,  
Forschungszentrum J\"ulich, D-52425 J\"ulich, Germany}
\affiliation{RWTH Aachen University, D-52056 Aachen, Germany}

\author{Jochen Gemmer}
\affiliation{Departement of Physics, University of Osnabr\"uck, D-49069 
Osnabr\"uck,
Germany}

\author{Robin Steinigeweg}\email{rsteinig@uos.de} 
\affiliation{Departement of Physics, University of Osnabr\"uck, D-49069 
Osnabr\"uck,
Germany}

\date{\today}

\begin{abstract}

For a class of typical states, the real-time and real-space dynamics of  
non-equilibrium density profiles has been recently studied for integrable 
models, i.e.\ the spin-1/2 XXZ chain [PRB \textbf{95}, 035155 (2017)] and the 
Fermi-Hubbard chain [PRE \textbf{96}, 020105 (2017)]. It has been found that 
the non-equilibrium dynamics agrees with linear response theory. 
Moreover, in the regime of strong interactions, clear signatures of diffusion 
have been observed. However, this diffusive behavior strongly depends on the 
choice of the initial state and disappears for untypical states without 
internal randomness. In the present work, we address the question whether or 
not the above findings persist for non-integrable models. As a first step, we 
study the spin-1/2 XXZ chain, where integrability can be broken due to an 
additional next-nearest neighbor interaction. Furthermore, we analyze the 
differences of typical and untypical initial states on the basis of their 
entanglement and their local density of states.

\end{abstract}


\maketitle

\section{Introduction}

Understanding the dynamics of quantum many-body systems constitutes a central
question in many areas of modern experimental and theoretical physics. While
this question has a long and fertile history, it has attracted continuously
increasing attention in the last decade \cite{polkovnikov2011, eisert2015}. This 
upsurge of interest is also related to the advent of novel materials and cold 
atomic gases \cite{bloch2008, langen2015}, the discovery of new states of 
matter such as many-body localized phases \cite{basko2006, nandkishore2015, 
abanin2017}, the invention of powerful numerical techniques such as 
density-matrix renormalization group \cite{schollwoeck2005, schollwoeck2011},
as well as the emergence of fresh key concepts, with typicality of pure states
\cite{goldstein2006, popescu2006, reimann2007, gemmer2003, sugiura2012, 
sugiura2013, elsayed2013, iitaka2003, iitaka2004, white2009, monnai2014, 
reimann2016} and eigenstate thermalization hypothesis \cite{deutsch1991, 
srednicki1994, rigol2008} as prime examples. Although clarifying the mere 
existence of equilibration and thermalization in isolated systems has seen 
substantial progress \cite{reimann2008, dalessio2016}, rigorously deriving the 
macroscopic phenomena of (exponential) relaxation and (diffusive) transport 
from truly microscopic principles is still a major challenge \cite{bonetto2000, 
buchanan2005}.

\begin{figure}[tb]
\centering
\includegraphics[width = 0.95\columnwidth]{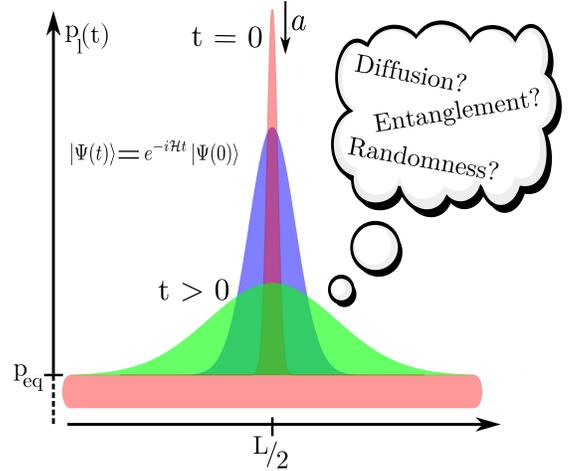}
\caption{(Color online) At time $t = 0$, the initial density profile exhibits a
central peak in the middle of the chain on top of a homogeneous many-particle 
background. The height of this peak can be controlled by an additional 
parameter $a>0$ (see appendix for details). In the present paper, the real-time 
broadening of such profiles is studied. In particular, we are interested in the 
role of entanglement and internal randomness of the pure state $\ket{\psi(0)}$.
} \label{PicInitState}
\end{figure}

In this context, two equally important questions are (i) the role of  
integrability and nonintegrability and (ii) the influence of the specific 
initial-state realization. On the one hand, integrable systems are 
characterized by a macroscopic number of (quasi)local conservation laws 
\cite{shastry1986, zotos1997, prosen2011, prosen2013, ilievski2016} and 
the overlap with these conserved quantities leads to unconventional 
equilibration and thermalization \cite{essler2016, vidmar2016, vasseur2016} and 
nondecaying currents \cite{zotos2004, heidrichmeisner2003, heidrichmeisner2007}. 
On the other hand, the overlap with one of the conserved quantities is not 
guaranteed for all parameters of the model, observables, and initial 
conditions. Therefore, integrability as such does not rule out the possibility 
of regular relaxation and transport processes. In fact, clear signatures of 
diffusion have been observed in both, the spin-$1/2$ Heisenberg chain above the 
isotropic point \cite{znidaric2011, steinigeweg2011_1, karrasch2014_2, 
steinigeweg2017_1}
and in the  Fermi-Hubbard model with strong onsite repulsion \cite{prosen2012, 
karrasch2017, steinigeweg2017_2}, at least in the limit of high temperatures. 
This remarkable observation suggests that nonintegrability, chaos, and 
ergodicity are no prerequisite for the existence of diffusion. However, it has 
also been demonstrated that the dynamics of integrable systems can strongly 
depend on details of the particular initial states chosen 
\cite{steinigeweg2017_1, steinigeweg2017_2}. Thus, an intriguing question is 
whether or not such a strong dependence can also appear in the case of 
integrability-breaking perturbations. In this case, another intriguing question 
is whether or not signatures of diffusion become more pronounced.

In this paper, we study these questions and focus, as a first step, on a 
nonintegrable version of the spin-$1/2$ Heisenberg model in one dimension. 
While integrability can be certainly broken in many different ways, we do so by 
taking into account an additional interaction between next-nearest neighbors.
For this model, we analyze the real-time and real-space dynamics of
magnetization as resulting for a convenient class of nonequilibrium initial 
states. These states have been introduced in \cite{steinigeweg2017_1}, are 
pure, and realize a sharp density peak on top of homogeneous many-particle 
background at any temperature, as illustrated in Fig.\ \ref{PicInitState}.  
Since this class of initial states allows for changing internal degrees of 
freedom without modifying the initial density profile, we are able to 
investigate whether and in how far such internal details influence the 
real-time broadening. Here, an useful concept is typicality of pure states. It 
implies in the case of internal randomness a dynamical behavior in agreement 
with the equilibrium correlation function and allows us to perform large-scale 
numerical simulations in the framework of linear response.

Summarizing our main results in a nutshell, we show that signatures of diffusion
are equally pronounced for the integrable and nonintegrable case. We further 
find in both cases a strong difference between the dynamics of typical 
states (with internal randomness) and untypical states (without any 
randomness). We further provide an explanation of this difference by a 
detailed analysis of entanglement and local density of states.

The rest of this paper is structured as follows. First, we introduce in 
Sec.\ \ref{ModelSec} the Heisenberg spin-1/2 chain with an 
integrability-breaking interaction between neighbors at next-nearest sites. 
Then, we discuss the framework in Sec.\ \ref{DRST_SEC} and give an overview 
over our observables and initial states, linear response, and diffusion. 
Afterward, we discuss in Sec.\ \ref{DQT} the concept of typicality 
and our numerical approach. Eventually, we present our results in Secs.\ 
\ref{RESULTSsec} and \ref{properties} and particularly analyze integrability 
vs.\ nonintegrability, typical vs.\ untypical states, as well as entanglement 
and local density of states. We finally close with a summary and conclusions in
Sec.\ \ref{CONCLsec} and provide additional information in the appendix.

\section{Model} \label{ModelSec}

The present paper studies the one-dimensional spin-$1/2$ XXZ chain, where 
the standard model is extended to incorporate also interactions between 
next-nearest neighbors. The Hamiltonian $\mathcal{H} = \mathcal{H}_{ 
\text{XXZ}} + \mathcal{H}^\prime$ with periodic boundary conditions reads 
\begin{gather}
\mathcal{H}_{\text{XXZ}} = J \sum_{l=1}^L \left( S_l^x S_{l+1}^x + S_l^y 
S_{l+1}^y + \Delta S_l^z S_{l+1}^z \right)\ , \label{Hamiltonian1} \\
\mathcal{H}^\prime = J \sum_{l=1}^L  \Delta^\prime S_l^{z} S_{l+2}^z\ ,
\label{Hamiltonian2}
\end{gather}
where $S_l^{i}$, $i \in \lbrace x, y, z\rbrace$ are spin-$1/2$ operators at site
$l$, $L$ is the total number of sites, and $J > 0$ is the antiferromagnetic 
exchange constant. Using the Jordan-Wigner transformation, $\mathcal{H}$ can be 
mapped to an one-dimensional model of spinless fermions with nearest and 
next-nearest neighbor interactions, where the strength of the interactions is 
set by $\Delta$ and $\Delta^\prime$, respectively. In the case $\Delta^\prime = 
0$, the model is integrable in terms of the Bethe Ansatz, with the energy 
current being exactly conserved \cite{zotos1997, kluemper2002}, whereas 
integrability is broken for any $\Delta^\prime \neq 0$.

\begin{figure}[tb]
\centering
\includegraphics[width=0.85\columnwidth]{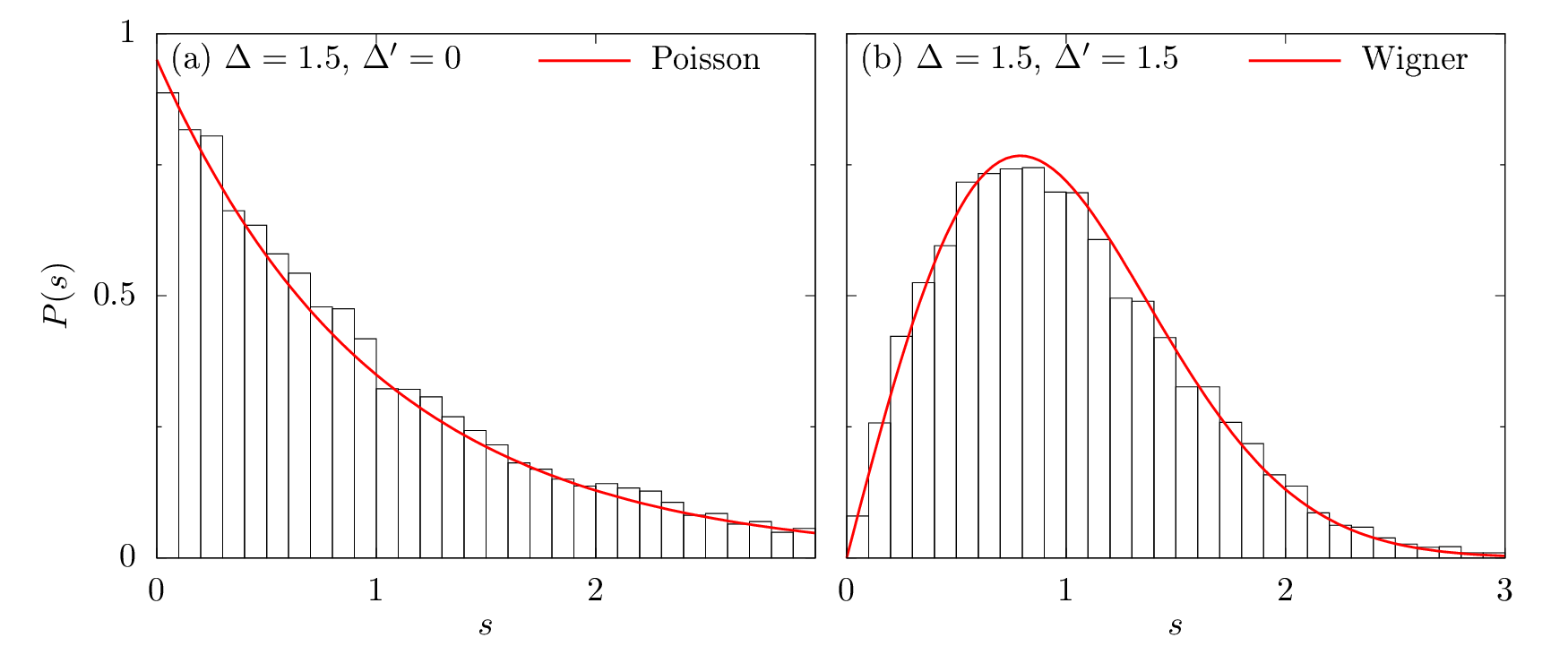}
\caption{(Color online) Level-spacing distribution $P(s)$ of the spin-$1/2$ XXZ
chain with $L = 20$, for a single symmetry subsector labeled by the quantum
numbers $S^z = 1$ and $k = 1$. In the integrable case $\Delta^\prime = 0$, the 
distribution is well described by a Poissonian, whereas for the non-integrable 
case $\Delta = \Delta^\prime = 1.5$, one observes Wigner statistics.}
\label{LevelStatistics_Pic}
\end{figure}

The difference between the integrable and the non-integrable model is also 
reflected in the level-spacing distribution $P(s)$, see Fig.\
\ref{LevelStatistics_Pic}. 
For $\Delta^\prime = 0$ [Fig.\ \ref{LevelStatistics_Pic} (a)], $P(s)$ exhibits
Poissonian behavior, in contrast to the non-integrable case $\Delta =
\Delta^\prime \neq 0$ [Fig.\ \ref{LevelStatistics_Pic} (b)], where $P(s)$ obeys
the quantum chaotic Wigner distribution. Note that a proper analysis of $P(s)$
requires an unfolding of the spectrum \cite{santos2004, santos2010}. 
Moreover, here we restrict ourselves to a single subsector of 
$\mathcal{H}$ with magnetization $S^z = 1$ and momentum $k = 1$, in order 
to eliminate all trivial symmetries. Note, however, that for the rest of this 
paper we always consider the full Hilbert space without any restriction.

\section{Framework} \label{DRST_SEC}

\subsection{Observables and initial states}

In this paper, the real-time dynamics of local occupation numbers
\begin{equation}
 n_l = S_l^z + \frac{1}{2}
\end{equation}
is studied. To this end, expectation values of the form
\begin{align}
 p_l(t)  = \text{Tr}[\rho(t)\ n_l] 
\end{align}
are evaluated, where $\rho(t)$ is the density matrix at time $t$,
\begin{equation}
\rho(t) = e^{-i\mathcal{H}t} \ket{\psi(0)}\bra{\psi(0)}
e^{i\mathcal{H}t}\ ,
\end{equation}
and $\ket{\psi(0)}$ is a pure state. The special class of (normalized)
\textit{non-equilibrium} initial states $\ket{\psi(0)}$ considered in this
paper are constructed as
\begin{equation} \label{initialState}
\ket{\psi(0)} \propto (n_{L/2} - a) \ket{\Phi}\ , \quad \ket{\Phi} = 
\sum_{k=1}^{2^L} c_k \ket{\varphi_k}\ ,
\end{equation}
where $c_k$ are complex coefficients and $a \geq 0$ is a real number. The states
$\ket{\varphi_k}$ denote the common eigenbasis of all $n_l$, i.e.\ the Ising 
basis. The operator $n_{L/2}$ acts as a projection onto all states with a
spin-up in the middle of the chain. In the case $a = 0$, we consequently have 
$p_{L/2}(0) = 1$ by construction. By choosing $a > 0$, it is however 
straightforward to adjust this initial amplitude. (For more details, see the
appendix). For the particular choice of all coefficients $c_k$
being the same in Eq.\ \eqref{initialState}, we moreover find $p_{l\neq L/2} =
p_{\text{eq}} = 1/2$. Thus, one ends up with an initial density profile which
has a central peak in the middle of the chain, on top of a homogeneous 
many-particle background, see Fig.\ \ref{PicInitState}. However, exactly the 
same density profile arises if real and imaginary part of the $c_k$ are 
randomly drawn from a Gaussian distribution with zero mean (according to the 
unitary invariant Haar measure \cite{bartsch2009, bartsch2011}).

Although not distinguishable at $t = 0$, it has been demonstrated              
\cite{steinigeweg2017_1} that the dynamics for times $t > 0$ can depend 
strongly on whether $\ket{\psi(0)}$ is a ``typical'' state with random
$c_k$ or an ``untypical'' state where all $c_k$ are the same. A central aim of 
the present paper is to understand the crucial differences between these two 
choices of initial states. To this end, the states are analyzed in terms of 
their local density of states, their internal randomness, as well as their 
entanglement. In this respect, it is important to note that, for all $c_k$ 
being the same, it is possible to write $\ket{\psi(0)}$ as a product state with 
a spin-up state $\ket{\uparrow}$ in the middle of the chain and 
a spin-up/spin-down superposition at all other sites,
\begin{equation}\label{ProdState}
\ket{\psi(0)}  \propto \dots (\ket{\uparrow} + \ket{\downarrow}) \otimes
\ket{\uparrow} \otimes (\ket{\uparrow} + \ket{\downarrow}) \dots\ .
\end{equation}
On the other hand, for completely or at least partially random coefficients 
$c_k$, such a full product structure is absent.

\subsection{Kubo Formula} \label{KuboSec}

Within the framework of linear response theory (LRT), transport coefficients
can be computed from current-current correlation functions 
\begin{equation}\label{CurCur}
 \langle j(t) j \rangle= \text{Tr} [j(t) j \rho_{\text{eq}}]\ ,
\end{equation}
which are evaluated within the canonical equilibrium ensemble $\rho_\text{eq} = 
e^{-\beta H}/{\cal Z}$ at inverse temperature $\beta = 1/T$ \cite{kubo1957, 
luttinger1964, mahan1990}, where $\mathcal{Z} = \text{Tr}[e^{-\beta {\cal H}}]$ 
is the partition function. The time argument has to be understood with respect 
to the Heisenberg picture.

In the present paper, our focus is on the transport of magnetization and the
operator $j$ therefore denotes the spin current. Since total magnetization $S^z
= \sum_l S_l^z$ is conserved, $[\mathcal{H},S^z] = 0$, the spin current $j$ is
well-defined and follows from the lattice continuity equation
\begin{align}
\frac{\text{d}}{\text{d}t} S_l^z = i [\mathcal{H},S_l^z] = j_{l-1} - j_l\ .
\end{align}
Thus, for the Hamiltonian $\mathcal{H}$, as defined in Eqs.\  
\eqref{Hamiltonian1}
and \eqref{Hamiltonian2}, $j$ takes on the well-known form
\begin{equation}
j = \sum_{l=1}^L j_l = J \sum_{l=1}^L (S_l^x S_{l+1}^y - S_l^y S_{l+1}^x)\ ,
\end{equation}
which is exactly conserved only in the case $\Delta = \Delta^\prime = 0$. In 
LRT, the connection between transport properties and current autocorrelations is
given by the {\it Kubo formula} which, in case of the spin current, can be 
written as
\begin{equation}\label{Kubo}
\sigma(\omega) = \frac{1-e^{-\beta\omega}}{\omega L}\int\limits_0^\infty  
e^{i\omega t}\ \langle j(t) j \rangle\ \text{d}t\ ,
\end{equation}
where $\sigma(\omega)$ is the conductivity at the inverse temperature 
$\beta$. Often, $\text{Re} \, \sigma(\omega)$ is decomposed into a $\delta$ 
function at $\omega = 0$ and a part for frequencies $\omega \neq 0$,
\begin{equation}
\text{Re}\ \sigma(\omega) = \bar{C} \delta(\omega) + \sigma_{\text{reg}}( 
\omega)\ , 
\end{equation}
where $\bar{C}$ is the so-called Drude weight \cite{zotos2004,  
heidrichmeisner2003, heidrichmeisner2007}. In fact, $\bar{C}$ can be directly 
related to the long-time limit of the current autocorrelation function $C(t)$ 
\cite{steinigeweg2009_1, karrasch2012, karrasch2013, steinigeweg2014, 
steinigeweg2015, karrasch2014_1},
\begin{equation}\label{DWEQ}
\bar{C} = \int_{t_1}^{t_2} \text{d}t\ \frac{C(t)}{t_2 - t_1}\ ,
\end{equation}
with $C(t) = \text{Re} \langle j(t) j \rangle/L$. Here, $t_1$ and $t_2$ are 
selected from a region where $C(t)$ has decayed to its long-time value 
$C(t\rightarrow \infty) \geq 0$. Thus, a nonzero Drude weight exists whenever 
the current is at least partially conserved and indicates ballistic transport 
\cite{zotos2004, heidrichmeisner2003, heidrichmeisner2007}. In cases where the 
Drude weight vanishes and transport is not ballistic in the thermodynamic 
limit, the dc conductivity $\sigma_{\text{dc}} = \sigma_{\text{reg}} (\omega 
\rightarrow 0)$ is of interest and follows from a zero-frequency 
Fourier transform of $C(t)$ \cite{prelovsek2004, steinigeweg2009_1, 
steinigeweg2015, karrasch2014_1, karrasch2014_2},
\begin{equation}\label{DCEq}
\sigma_{\text{dc}} = \beta\int_0^{t_{\text{max}}} \text{d}t\ C(t)\ . 
\end{equation}
Since the Drude weight $\bar{C}$ will always be nonzero for a finite system,  
the integral in Eq.\ \eqref{DCEq} diverges in the limit $t_{\text{max}} 
\rightarrow \infty$ \cite{steinigeweg2009_1}. Therefore the cutoff time 
$t_{\text{max}} < \infty$ is chosen to be finite, but long enough to ensure 
that $\sigma_{\text{dc}}$ is effectively independent of the particular choice 
of $t_{\text{max}}$. Note that there exist different definitions for $\bar{C}$ 
and $\sigma_{ \text{dc}}$ in the literature, 
with additional prefactors $\pi$, $2\pi$ and $\beta$.

\subsection{Diffusion}

As discussed in Sec.\ \ref{KuboSec}, a finite Drude weight immediately implies
ballistic transport. However, a vanishing Drude weight not necessarily leads to
diffusive behavior. In this subsection, we therefore summarize the conditions 
for diffusion.

Defined on a discrete lattice, the dynamics of some density (here magnetization 
density) $p_l$ is said to be diffusive if it fulfills a diffusion equation of 
the form \cite{michel2005, steinigeweg2007}
\begin{equation} \label{DiffEQ}
\frac{\text{d}}{\text{d}t}\ p_l(t) = D \left[ p_{l-1}(t) - 2 p_l(t) + p_{l+1}(t)
\right]\ ,  
\end{equation}
where $D$ is the \textit{time-independent} diffusion constant. For this 
equation, one finds a specific solution for the time and site dependence of 
$p_l(t)$,
\begin{equation}
p_l(t) - p_{\text{eq}} = \frac{1}{2} \exp(-2 D t) \, {\cal B}_{l-L/2}(2 D t)\ ,
\end{equation}
with ${\cal B}_l(t)$ being the modified Bessel function of the first kind. This
lattice solution can be well approximated by the corresponding continuum 
solution
\begin{equation}\label{GaussEq}
p_l(t) - p_{\text{eq}} = \frac{1}{2} \frac{1}{\sqrt{2\pi} \Sigma(t)} \exp 
\left[ - \frac{(l - L/2)^2}{2\Sigma^2(t) } \right]\ ,
\end{equation}
where the spatial variance is given by 
\begin{equation}\label{SigmaEqEQ}
\Sigma^2(t) = 2 D t\ .
\end{equation}
Note that, in the limit $\Sigma(t \rightarrow 0)$, Eq.\ \eqref{GaussEq} becomes
a $\delta$ function located at lattice site $l = L/2$, which coincides with our
initial density profile.

Generally, the spatial variance $\Sigma^2(t)$ of an arbitrary distribution is 
given by
\begin{equation}\label{Sigma_Eq}
\Sigma^2(t) = \sum_{l=1}^L l^2\ \delta p_l(t) - \left[ \sum_{l=1}^L l\ \delta 
p_l(t)\right]^2\ , 
\end{equation}
where $\delta p_l(t) \propto [p_l(t) - p_{\text{eq}}]$ and $\sum_{l=1}^L \delta
p_l(t) = 1$. Thus, in the case of diffusive transport, the variances from Eqs.\
\eqref{SigmaEqEQ} and \eqref{Sigma_Eq} exactly coincide with each other, and 
our non-equilibrium dynamics should be described by Gaussians as
given in Eq.\ \eqref{GaussEq}.

However, a time-independent diffusion constant, and the existence of diffusion 
as such, is questionable in view of unitary Schr\"odinger dynamics
\cite{michel2005}. Moreover, as we will also see during the discussion of our
results, it might not always be appropriate to draw conclusions only on the
basis of the real-space data. Therefore, we here introduce an useful
scheme: A Fourier transform of the diffusion equation in Eq.\ \eqref{DiffEQ}
yields
\begin{align}\label{DiffMom}
\frac{\text{d}}{\text{d}t}\ p_q(t) = -2(1-\cos q) D_q(t) p_q(t)\ ,
\end{align}
where we additionally allow for a time- and momentum-dependent $D_q(t)$, and 
momentum $q$ takes on the values $q = 2 \pi k/L$ with $k = 0, 1, \ldots,
L-1$. Rearranging Eq.\ \eqref{DiffMom} and using the abbreviation $\tilde{q}^2 =
2(1- \cos q)$ then gives the \textit{generalized diffusion coefficient}
\cite{steinigeweg2011_1}
\begin{equation}\label{GenDiffC}
D_q(t) = \frac{\text{d}/\text{d}t\ p_q(t)}{-\tilde{q}^2\ p_q(t)}\ .
\end{equation}
In the case of diffusive transport, the behavior of $D_q(t)$ can be 
qualitatively understood as follows. On the one hand, $D_q(t) \propto t$ always
increases linearly for sufficiently short times \cite{steinigeweg2011_1}. On 
the other hand, above the mean free time $\tau$ and above the mean free path
$\lambda$, i.e., $t > \tau$ and $\pi/q > \lambda$, $D_q(t)$ eventually turns 
into a plateau with $D_q(t) \approx \text{const}$, which marks the hydrodynamic
regime.

Eventually, it is also instructive to connect $D_q(t)$ to linear response 
theory. {\it Assuming} $p_q(t) \propto \text{Re} \, \langle S_q^z(t) S_{-q}^z 
\rangle$, where $S_q^z = \sum_l e^{iql} S_l^z/\sqrt{L}$, it follows in the limit 
$q \rightarrow 0$ that \cite{steinigeweg2011_1}
\begin{equation}\label{DiffK_Eq}
D(t) = \frac{1}{\chi} \int_0^t \text{d}t^\prime\ C(t^\prime)\ , 
\end{equation}
where the static susceptibility is $\chi = 1/4$ in the limit $\beta \rightarrow 
0$. Under the above {\it assumption}, $D(t)$ is also related to the time 
derivative of the spatial variance \cite{steinigeweg2009_2, yan2015, luitz2017, 
karrasch2017},
\begin{equation}\label{LR_Sigma_EQ}
\frac{\text{d}}{\text{d}t} \Sigma^2(t) = 2 D(t)\ . 
\end{equation}
The time dependence of $D(t)$ can be summarized as follows. For the 
non-interacting case $\Delta = \Delta^\prime = 0$, we have $[\mathcal{H},j] = 
0$, leading to $D(t) \propto t$ such that $\Sigma^2(t) \propto t^2$ scales 
ballistically for all t. Such ballistic behavior is also known to occur for 
partial current conservation at $\Delta < 1$ and $\Delta' = 0$ 
\cite{shastry1990, castella1995, narozhny1998, zotos1999, heidrichmeisner2003, 
fujimoto2003, benz2005, prosen2011, prosen2013, herbrych2011, karrasch2012, 
karrasch2013, steinigeweg2014, steinigeweg2015, ilievski2017}. In the case of 
diffusive transport, $D(t) = \text{const}$ and $\Sigma(t) \propto t$. Moreover, 
a process is called superdiffusive if $\Sigma(t) \propto t^\alpha$ with $\alpha 
\in ]1,2[$ and subdiffusive for $\alpha \in ]0,1[$. However, it is important to 
note that $D(t)$ yields no information beyond the mere width of density 
profiles.

\section{Dynamical Quantum Typicality} \label{DQT}

\subsection{Current-current correlations}

The concept of typicality \cite{gemmer2003, goldstein2006, popescu2006, 
reimann2007, sugiura2012, sugiura2013, elsayed2013, 
iitaka2003, iitaka2004, white2009, monnai2014, reimann2016} states 
that a single pure state can have the same ``properties'' as the full 
statistical ensemble. Remarkably, this concept does not require eigenstate 
thermalization \cite{deutsch1991, srednicki1994, rigol2008} and also applies to 
the {\it dynamics} of expectation values. In particular, dynamical quantum 
typicality (DQT) has turned out to be a powerful method for the accurate 
calculation of real-time current correlation functions in huge Hilbert spaces 
\cite{elsayed2013, steinigeweg2014, steinigeweg2015, steinigeweg2016}.

The main idea is to replace the trace $\text{Tr}[ \bullet ]$ in Eq.\ 
\eqref{CurCur} by a single scalar product $\bra{\Phi} \bullet \ket{\Phi}$, 
where $\ket{\Phi}$ is a pure state, randomly drawn from the full Hilbert space
according to the unitary invariant Haar measure \cite{bartsch2009, 
bartsch2011}. The current autocorrelation function can then be written 
as \cite{elsayed2013, steinigeweg2014, steinigeweg2015, steinigeweg2016}
\begin{equation}
C(t) = \frac{\text{Re} \bra{\Phi} j(t)\ j\ e^{-\beta \mathcal{H}} \ket{\Phi}}{L 
\bra{\Phi} e^{-\beta \mathcal{H}}\ket{\Phi}} + \epsilon(\ket{\Phi})\  
\label{TypiEQ}
\end{equation}
or, equivalently, as
\begin{equation}
C(t) = \frac{\text{Re} \bra{\phi(t)} j \ket{\varphi(t)}}{L\braket{\phi(0)|
\phi(0)}} + \epsilon(\ket{\Phi})\ ,
\end{equation}
where we have introduced the two auxiliary pure states 
\begin{align}
&\ket{\phi(t)} = e^{-i\mathcal{H}t} e^{-\beta {\cal H}/2} \ket{\Phi}\ ,
\label{State1} \\
&\ket{\varphi(t)} = e^{-i\mathcal{H}t} j\ e^{-\beta {\cal H}/2}
\ket{\Phi}\ , \label{State2}
\end{align}
which only differ by the additional current operator in Eq.\ \eqref{State2}. It
is important to note that the error in Eq.\ \eqref{TypiEQ} scales as $\epsilon
\propto 1/\sqrt{d}$ for $\beta \rightarrow 0$, with $d=2^L$ being the dimension 
of the Hilbert space. Thus, for the large system sizes we are interested in, 
this error is negligibly small and the typicality approximation can be regarded 
as practically exact. Furthermore, the time dependence, e.g.\ of 
$\ket{\phi(t)}$, can be conveniently evaluated by iteratively solving the 
real-time Schr\"odinger equation (see Sec.\ \ref{NUMsec}).

\subsection{Density-density correlations}

Concerning the dynamics of local occupation numbers, we can perform the 
following calculation \cite{steinigeweg2017_1, steinigeweg2017_2}. We start 
from an equilibrium correlation function in the limit $\beta \to 0$,
\begin{align}
\mathcal{C}_l(t) = 2 \langle n_{L/2}\ n_l(t) \rangle &= 2 \frac{\text{Tr}[
n_{L/2}\ n_l(t)]}{2^L} \\
&= 2\frac{\text{Tr}[n_{L/2}\ n_l(t)\ n_{L/2}]}{2^L}\ ,
\end{align}
where the cyclic invariance of the trace and the projection property $n_{L/2}^2
= n_{L/2}$ has been exploited. According to typicality, also this expression 
can be rewritten using a randomly drawn pure state 
$\ket{\Phi}$, 
\begin{align}
\mathcal{C}_l(t) &= 2 \frac{\bra{\Phi}n_{L/2}\ n_l(t)\ n_{L/2} \ket{\Phi}}
{\braket{\Phi|\Phi}} + \epsilon(\ket{\Phi}) \\
&= \frac{\bra{\psi} e^{i\mathcal{H}t}\ n_l\ e^{-i\mathcal{H}t}
\ket{\psi}}{\braket{\psi|\psi}}\ ,
\end{align}
where we have used the definition of our initial state in Eq.\
\eqref{initialState} and $\braket{\psi|\psi} = \braket{\Phi| \Phi}/2$. 
Moreover, we have dropped the error $\epsilon$ for clarity. Since 
$\ket{\psi(t)} = e^{-i\mathcal{H}t} \ket{\psi}$, we finally find 
\begin{equation}
\mathcal{C}_l(t) = \frac{\bra{\psi(t)} n_l \ket{\psi(t)}}
{\braket{\psi(0)|\psi(0)}} = p_l(t)\ .
\end{equation}
Thus, it follows that, although the initial states in Eq.\ \eqref{initialState},
have to be considered as \textit{far from equilibrium}, the resulting
non-equilibrium dynamics is directly related to an equilibrium correlation
function. 

\subsection{Forward propagation of pure states} \label{NUMsec}

Using exact diagonalization (ED), it is possible to compute the time evolution
of a pure state via 
\begin{equation}
\ket{\psi(t)} = \sum_n e^{iE_nt} c_n \ket{n}\ ,
\end{equation}
where $\ket{n}$ are eigenvectors of the Hamiltonian with corresponding 
eigenvalues $E_n$, and $c_n = \braket{n|\psi(0)}$ denotes the overlap of 
$\ket{n}$ and $\ket{\psi(0)}$. However, the exponential growth of the Hilbert 
space represents a natural limitation of ED. Usually, this growth is at least 
partially compensated by exploiting the symmetries of the Hamiltonian. To 
repeat, the Hamiltonian $\mathcal{H}$ in Eqs.\ \eqref{Hamiltonian1} and 
\eqref{Hamiltonian2} conserves total magnetization $S^z = \sum_l S_l^z$. 
Moreover, it is invariant under translation by one lattice site and crystal 
momentum $k$ becomes a good quantum number. Thus, it is in principle possible 
to divide the Hilbert space into subspaces, classified by $S^z$ and $k$. 
However, since the operator $n_{L/2}$ in the definition \eqref{initialState} of 
the initial states does not respect translational invariance, it becomes less 
profitable to use this symmetry for our calculations. In any case, ED is 
limited to systems with a maximum of $L \sim 20$ sites.

Therefore, we proceed differently in the present paper and rely on a forward
propagation of $\ket{\psi(t)}$ in real time. Such a propagation can be
done by means of a fourth-order Runge-Kutta (RK4) scheme \cite{elsayed2013, 
steinigeweg2014, steinigeweg2015, steinigeweg2016} or by more sophisticated 
methods such as Chebyshev polynomials \cite{weisse2006, dobrovitski2003} 
or Trotter decompositions \cite{steinigeweg2017_1, steinigeweg2017_2, 
deReadt2006}. Using these methods, no diagonalization of $\mathcal{H}$ is 
needed and, since $\mathcal{H}$ is usually relatively sparse, the matrix-vector 
multiplications can be implemented very memory-efficient. In this paper, we use 
a RK4 method for chains up to $L \leq 26$ sites. For longer chains, we employ a 
Trotter product formula which allows us to treat systems with as many as $L = 
36$ spins. For this $L$, the largest subsector with $S^z = 0$ has dimension $d 
\approx 10^{10}$ and is several orders of magnitude larger than the matrices 
treatable by state-of-the-art ED.

\section{Dynamics of Typical and Untypical States} \label{RESULTSsec}

We now present our numerical results. As a first step in Sec.\ \ref{CurSEc}, we 
study current autocorrelations and Drude weights, i.e.\ results obtained within 
the framework of LRT. These results will be useful in the discussion of the 
non-equilibrium dynamics in the subsequent Sec.\ \ref{ResTyp}.

\subsection{Current autocorrelations and Drude weights} \label{CurSEc}

\begin{figure}[tb]
\includegraphics[width=0.85\columnwidth]{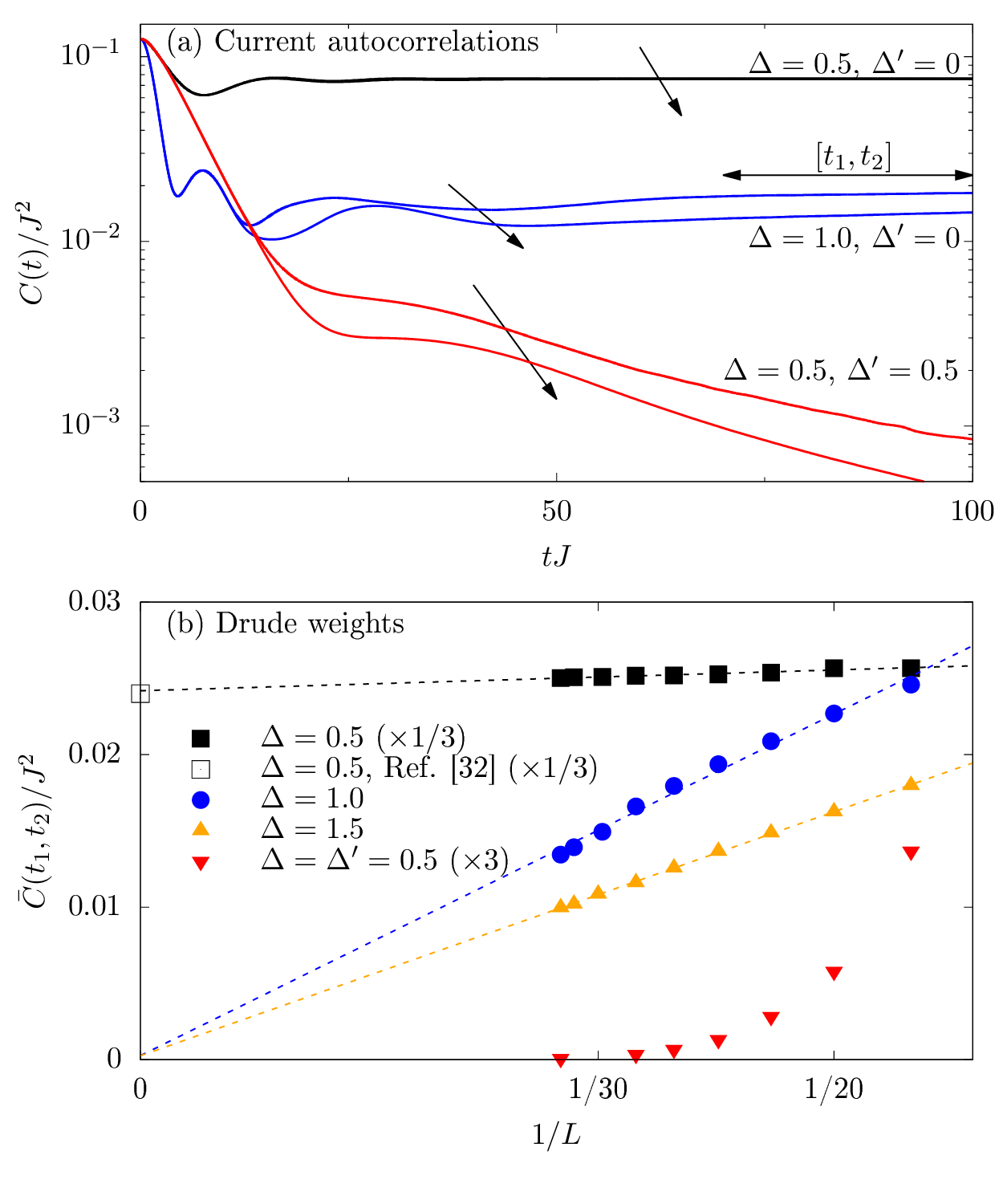}
\caption{(Color online) (a) Current autocorrelation function $C(t)$ up to times
$tJ = 100$ and for systems with $L = 26$ and $33$ sites (arrows). Data is shown 
for the integrable case $\Delta^\prime = 0$ with $\Delta = 0.5$ and $\Delta
= 1$ as well as for the non-integrable case $\Delta = \Delta^\prime = 0.5$. (b) 
Finite-size scaling of the Drude weight $\bar{C}$ for selected values of 
$\Delta$ and $\Delta^\prime$. For the integrable cases $\Delta^\prime = 0$, the 
data is obtained according to Eq.\ \eqref{DWEQ} and from the finite time 
interval $[t_1 J, t_2 J] = [70,100]$, cf.\ Fig.\ \ref{CurCur_DW} (a), whereas 
for $\Delta = \Delta^\prime = 0.5$, the interval $[t_1 J, t_2 J] = [250, 300]$ 
is chosen. The dashed lines are linear fits to the data. In the case $\Delta = 
0.5$, $\Delta = 0$ we additionally show an analytic bound for $\bar{C}$ 
\cite{prosen2011, prosen2013}. Note that $L = 33$ data for the integrable cases 
have been taken from Ref.\ \cite{steinigeweg2014}.
}
\label{CurCur_DW}
\end{figure}

According to Eq.\ \eqref{DWEQ}, the Drude weight $\bar{C}$ is related to the
long-time limit of the current-current correlation function $C(t)$. Since
$\bar{C} > 0$ for \textit{finite} systems, a careful finite-size scaling needs
to be performed, in order to draw reliable conclusions on $\bar{C}$ in the  
thermodynamic limit. Therefore, in Fig.\ \ref{CurCur_DW} (a), $C(t)$ is shown 
for different choices of $\Delta$ and $\Delta^\prime$ and for various chain 
lengths $L$.

While it is certainly convenient to start our discussion with the integrable 
model, i.e.\ $\Delta^\prime = 0$, we should stress that corresponding
results and a detailed discussion can be found already in Ref.\ 
\cite{steinigeweg2014}. For $\Delta^\prime= 0$ and at the isotropic point
$\Delta = 1$, one observes that, after an initial decay, $C(t)$ reaches an
approximately constant long-time value for times $tJ \gtrsim 50$.
Moreover, this long-time value decreases for increasing system size. On the
contrary, for $\Delta^\prime = 0$ and $\Delta = 0.5$, a significant
dependence of $C(t)$ and its long-time value on $L$ is not visible. 
Most important, however, in the case of a non-zero next-nearest neighbor
interaction $\Delta^\prime = 0.5$, $C(t)$ decays to substantially smaller
values. In fact, even at times $tJ = 300$ (not shown), $C(t)$ has not yet
reached its stationary value.

In Fig.\ \ref{CurCur_DW} (b), we show a finite-size scaling of the Drude weight
$\bar{C}$. For the integrable model, the data is obtained according to Eq.\
\eqref{DWEQ} and from the finite time interval $[t_1 J, t_2 J] = [70,100]$, as 
indicated in Fig.\ \ref{CurCur_DW} (a). Linear extrapolations of the data 
towards the thermodynamic limit are also depicted. In the case $\Delta = 0.5$, 
one observes that the Drude weight converges towards a finite value $\bar{C} > 
0$, in quantitative agreement with analytical results \cite{prosen2011, 
prosen2013}. For the case $\Delta \geq 1$, the linear fit clearly suggest a 
vanishing Drude weight $\bar{C} = 0$ for $L\rightarrow \infty$. For the 
non-integrable model $\Delta = \Delta^\prime = 0.5$, $\bar{C}$ is extracted 
from the interval $[t_1 J, t_2 J] = [250,300]$. As mentioned, $C(t)$ has not 
completely decayed even at these long times such that the data has to be 
understood as an upper bound for $\bar{C}$. Apparently, this upper bound
decreases faster than a power law with increasing $L$ and is most likely
expected to vanish for $L \rightarrow \infty$, as expected for non-integrable
systems \cite{heidrichmeisner2003, heidrichmeisner2007}.

\subsection{Real-space dynamics of typical states} \label{ResTyp}

\begin{figure}[t]
\includegraphics[width=0.85\columnwidth]{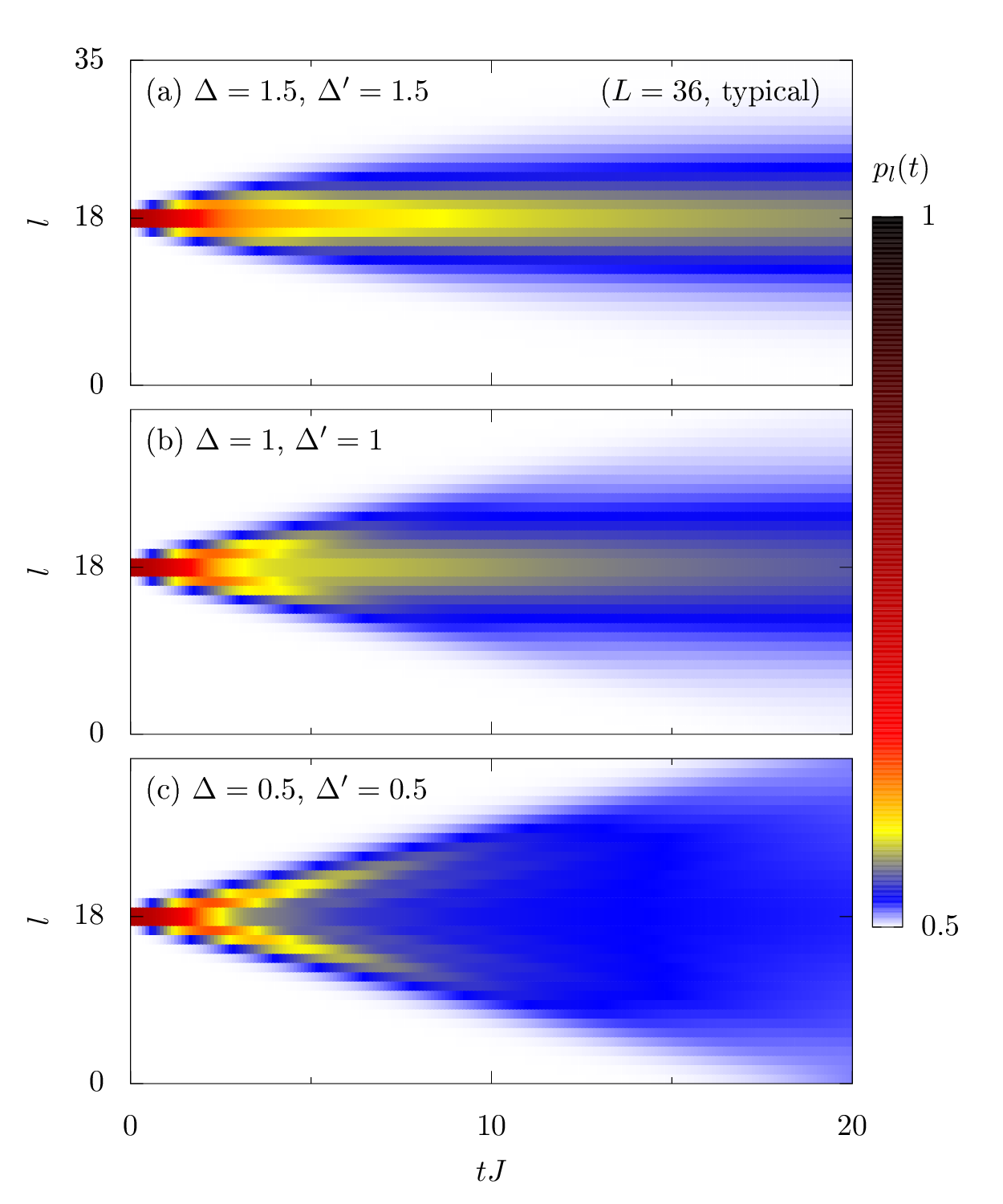}
\caption{(Color online) Time-space density plot of occupation numbers $p_l(t)$
for a \textit{typical} initial state $\ket{\psi(0)}$ in the XXZ spin-$1/2$ 
chain with $L = 36$ sites and different anisotropies $\Delta = \Delta^\prime = 
1.5$, $1$, and $0.5$ [(a) - (c)].}
\label{DichteC_Ty_Pic}
\end{figure}

To start the discussion of non-equilibrium dynamics in real space and time, we
first consider \textit{typical} initial states. In Fig.\ \ref{DichteC_Ty_Pic},
a time-space density plot of occupation numbers $p_l(t)$ is shown for a chain
with $L = 36$ sites and different anisotropies $\Delta = \Delta^\prime = 1.5$,
$1$, $0.5$, up to times $t J = 20$. For all parameters shown, one observes that 
the sharp initial peak broadens monotonically with time. In the case of weak 
interactions [Fig.\ \ref{DichteC_Ty_Pic} (c)], this broadening is still linear 
due to a long mean free time $\tau = \mathcal{O}(10)$. This can be also 
understood with respect to the current autocorrelation [see Fig.\ 
\ref{CurCur_DW} (a)], which is not fully decayed at this time scale. On the 
other hand, for larger anisotropies, the broadening of the density profiles is 
non-linear and significantly slower, which can be explained by the increased 
scattering of particles.

\begin{figure}[b]
\centering
\includegraphics[width=0.9\columnwidth]{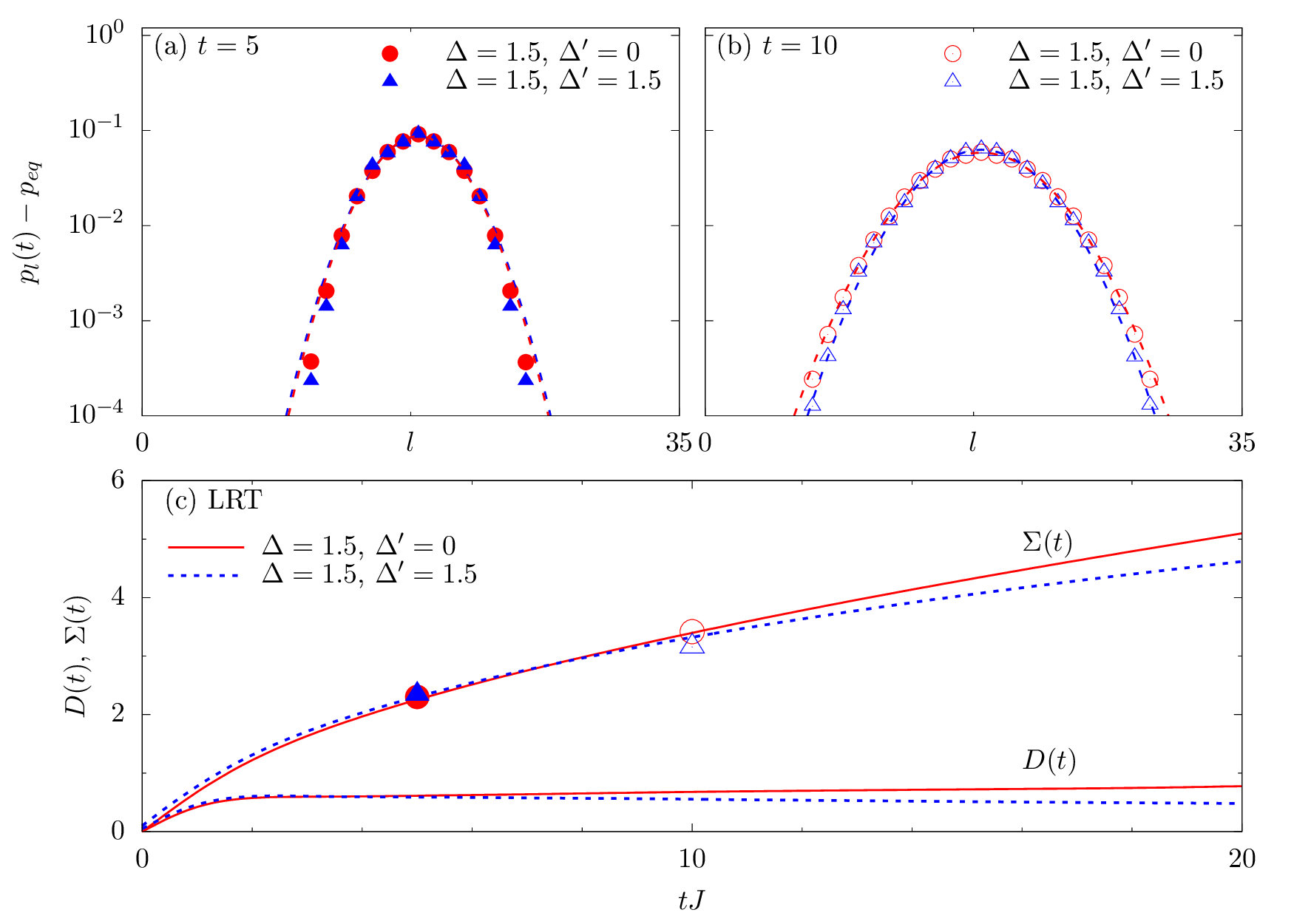}
\caption{(Color online) (a) and (b) Density profile $p_l(t)$ with respect to  
site $l$ at fixed times $tJ = 5$ and $10$ for $\Delta = 1.5$, $\Delta^\prime = 
0$ as well as $\Delta = \Delta^\prime = 1.5$ shown in a semi-log plot. The 
difference between the integrable and non-integrable model are remarkably small 
and the data is well described by Gaussian fits over several orders of 
magnitude. (c) Time-dependent diffusion coefficient $D(t)$ and profile width 
$\Sigma(t)$ according to LRT for $L = 34$ ($\Delta' = 0$) 
\cite{steinigeweg2015} and $L = 36$ ($\Delta' = 1.5$). For comparison, the 
symbols represent the width $\Sigma(t)$ of the non-equilibrium data in (a) and 
(b) and are in convincing agreement with LRT.} \label{Density_Times_15}
\end{figure}

For a more detailed analysis, Fig.\ \ref{Density_Times_15} (a) shows the 
density profile $p_l(t)$ for fixed times $t J = 5$ and $10$ in a semi-log plot, 
both for the integrable case with $\Delta = 1.5$, $\Delta^\prime = 0$ and the 
non-integrable case with $\Delta = \Delta^\prime = 1.5$. One observes that the 
data is remarkably well described by Gaussians over several orders of 
magnitude. Moreover, there are no significant differences between the integrable 
and the non-integrable model visible. Thus, we conclude that, for a large 
anisotropy $\Delta = 1.5$, the dynamics of our typical initial state is 
basically unaffected by the strong additional next-nearest neighbor interaction, 
which can be also explained analytically on the basis of projection operator 
techniques \cite{steinigeweg2011_2}.

In Fig.\ \ref{Density_Times_15} (b), we additionally compare the non-equilibrium
dynamics to results from LRT. To this end, the time-dependent diffusion 
coefficient $D(t)$ and the corresponding width $\Sigma(t)$ [see Eqs.\ 
\eqref{DiffK_Eq} and \eqref{LR_Sigma_EQ}] are shown for $L = 36$ sites. These 
LRT results are compared to the values of $\Sigma(t)$ according to Eq.\ 
\eqref{Sigma_Eq}, i.e., as directly extracted from the density profiles in 
Fig.\ \ref{Density_Times_15} (a). Overall, we find a convincing agreement 
between the non-equilibrium dynamics and LRT. Most importantly, however, one 
observes $D(t) \approx \text{const}$ at the time scales depicted 
\cite{steinigeweg2009_1, karrasch2014_2}. Thus, $\Sigma(t) \propto \sqrt{t}$, 
both for $\Delta^\prime = 0$ and $\Delta^\prime \neq 0$. This scaling as well 
as the Gaussian form of the density profiles clearly indicate diffusive 
transport in this parameter regime, irrespective of the model being integrable 
or non-integrable. This is a central result of our paper.

\begin{figure}[tb]
\centering
\includegraphics[width=0.9\columnwidth]{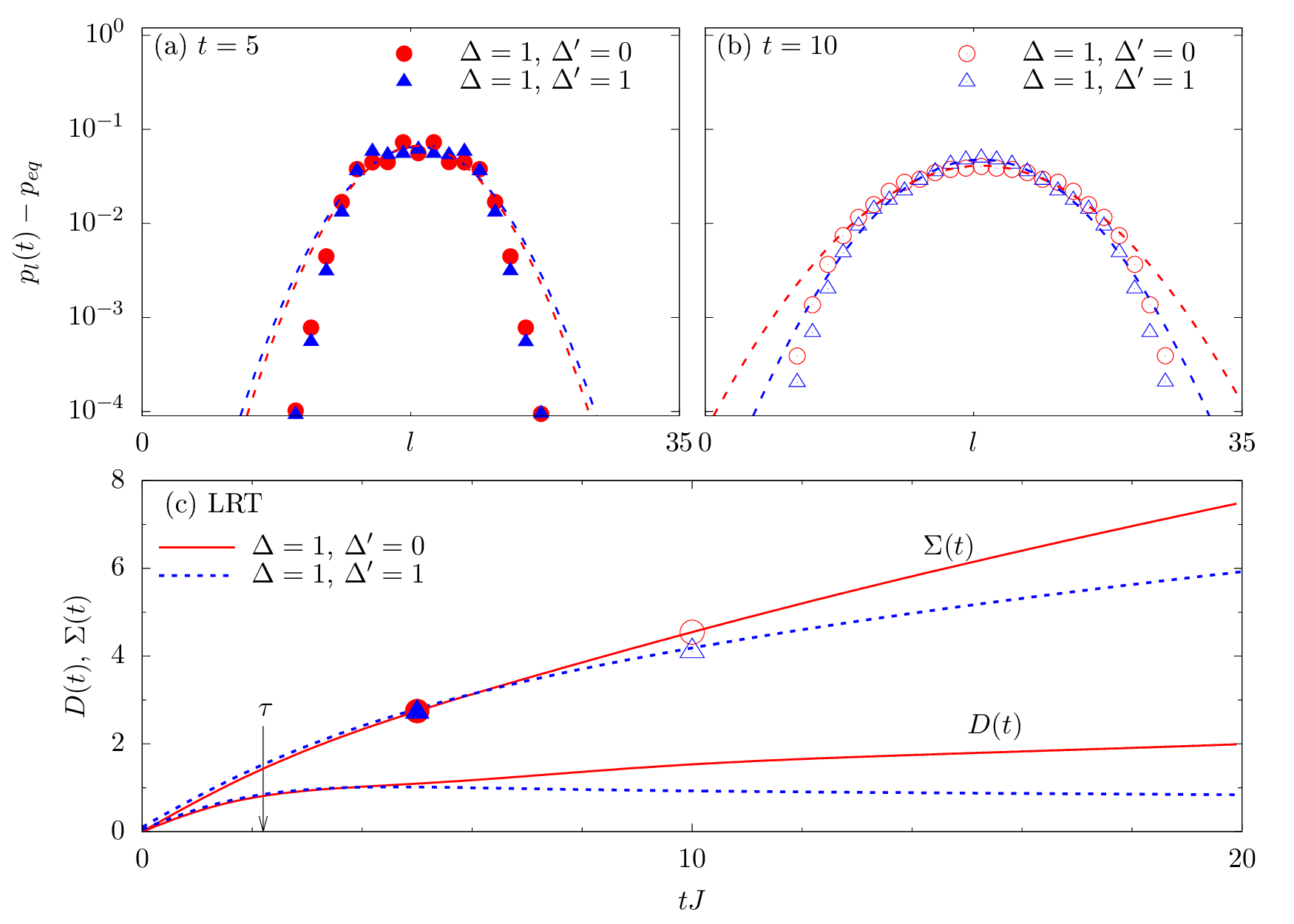}
\caption{(Color online) (a) Density profile $p_l(t)$ with respect to site $l$ 
at fixed times $tJ = 5$, $10$ for $\Delta = 1$, $\Delta^\prime = 0$ and $\Delta 
= \Delta^\prime = 1$, shown in a semi-log plot. (b) Time-dependent diffusion 
coefficient $D(t)$ and profile width $\Sigma(t)$ according to LRT for $L = 34$ 
($\Delta' = 0$) \cite{steinigeweg2014} and $L = 36$ ($\Delta' = 1$). The 
symbols represent the width $\Sigma(t)$ of the non-equilibrium data in (a). 
$\tau$ approximately marks the mean free time.} \label{Density_Times_1}
\end{figure}

Next, let us discuss the case of smaller $\Delta$ and $\Delta^\prime$ in more 
detail. Completely analogous to Fig.\ \ref{Density_Times_15}, the density 
profiles $p_l(t)$ for $\Delta = 1$, $\Delta^\prime = 0$ and $\Delta = 
\Delta^\prime = 1$ are shown in Fig.\ \ref{Density_Times_1} (a) for fixed times 
$t J = 5$ and $10$.  Compared to the previous case of larger anisotropies, we 
observe that it is not possible anymore to describe the density profiles by 
Gaussian fits, both for the integrable and the non-integrable model. 
Furthermore, in contrast to the case of larger anisotropies, the time 
dependence of $D(t)$ and $\Sigma(t)$ exhibits significant differences between
$\Delta' = 0$ and $\Delta' \neq 0$. On the one hand, the nonconstant $D(t)$ in 
the integrable case is clearly inconsistent with diffusion but rather suggests 
superdiffusive behavior \cite{znidaric2011, steinigeweg2012, ljubotina2017}, see 
also \cite{khait2016}. In contrast, at low temperatures, signatures of 
diffusive behavior have been reported \cite{sirker2009, sirker2011, 
grossjohann2010}. On the other hand, for $\Delta^\prime \neq 0$, one observes 
$D(t) \approx \text{const}$ as well as $\Sigma(t) \propto \sqrt{t}$. However, 
due to the non-Gaussian density profiles in Fig.\ \ref{Density_Times_1} (a), 
one might argue that the possibility of diffusion is still ruled out. It should 
be noted, however, that for times below the mean free time $\tau J \approx 2$ 
one finds $D(t) \propto t$ [see Fig.\ \ref{Density_Times_1} (b)] and only for 
times $t > \tau$ the diffusion coefficient $D(t)$ turns into a constant 
plateau. Thus, at short times, the sharp initial density profile broadens 
ballistically. Consequently, even if there exists diffusive behavior at longer  
time scales, one generally cannot expect clean Gaussian profiles but rather a 
superposition of such Gaussians.

\subsection{Momentum-space dynamics of typical states}

Due to the above reasoning, it is sometimes not sufficient to draw conclusions 
on diffusive or non-diffusive behavior only on the basis of the real-space 
data, with single-site resolution below the mean free path. Consequently, we 
proceed also in a different way and analyze the generalized diffusion 
coefficient $D_q(t)$, as introduced in Eq.\ \eqref{GenDiffC}.

\begin{figure}[b]
\centering
\includegraphics[width=1\columnwidth]{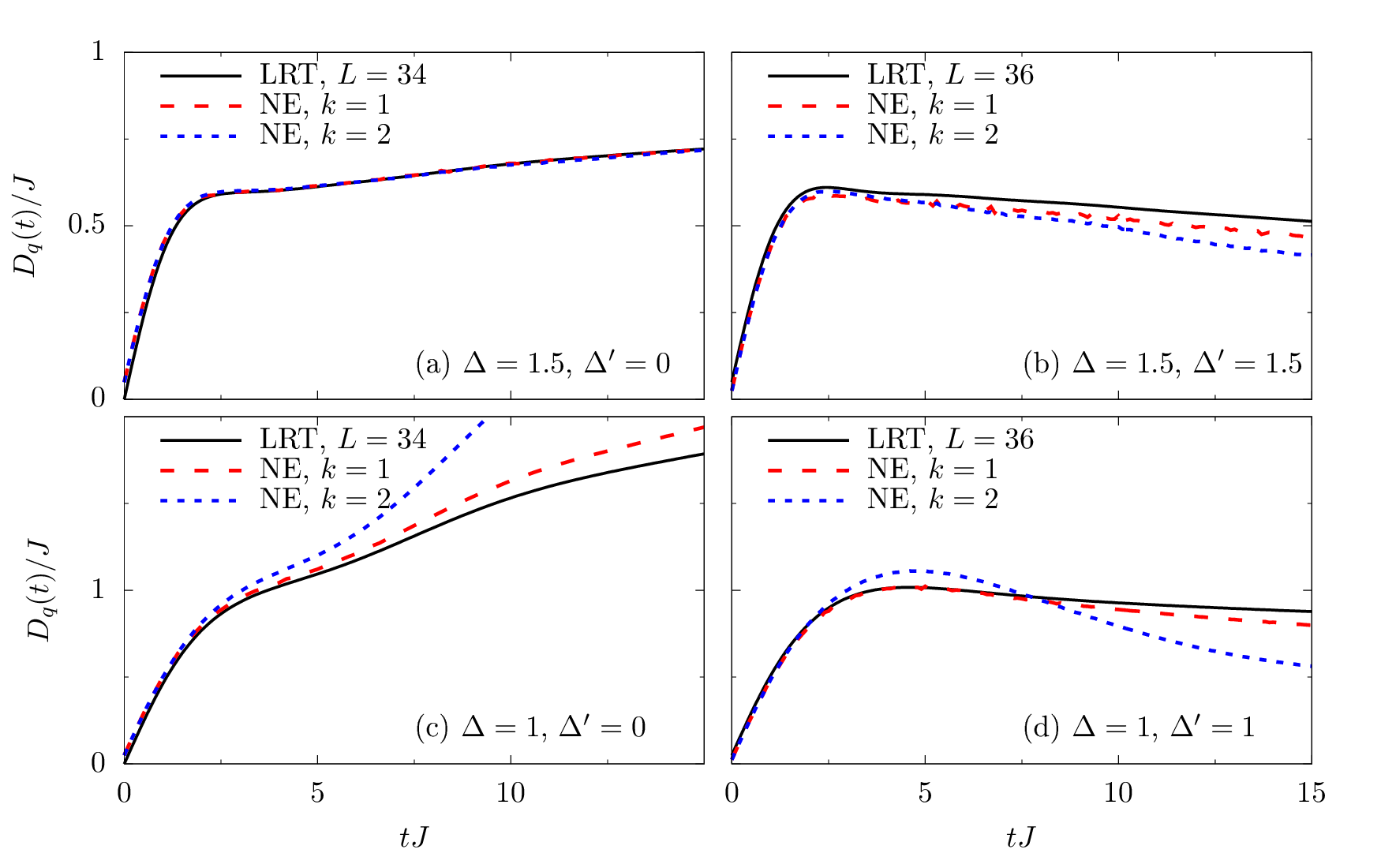}
\caption{(Color online) Generalized diffusion coefficient $D_q(t)$, obtained 
from the non-equilibrium density profiles according to Eq.\ \eqref{GenDiffC} 
for momenta $q/(2\pi/L) = k$ and $L = 36$. As a comparison, 
$D_{q=0}(t)$ according to LRT is shown, for $L = 34$ ($\Delta' = 0$) 
\cite{steinigeweg2015} and $L = 36$ ($\Delta' \neq 0$). Other parameters: (a) 
$\Delta = 1.5$, $\Delta^\prime = 0$; (b) $\Delta = \Delta^\prime = 1.5$; (c) 
$\Delta = 1$, $\Delta^\prime = 0$; (d) $\Delta = 1$, $\Delta^\prime = 1$.} 
\label{Dq_All}
\end{figure}

In Figs.\ \ref{Dq_All} (a) and (b), the generalized diffusion coefficient 
$D_q(t)$ is shown for large anisotropies $\Delta = 1.5$, $\Delta^\prime = 0$ 
and $\Delta = \Delta^\prime = 1.5$. Non-equilibrium results at momentum 
$q/(2\pi/L) = 1$ and $2$ are compared to LRT for $q = 0$, up to times $t J = 
15$. Overall, the integrable model in Fig.\ \ref{Dq_All} (a) and the 
non-integrable model in Fig.\ \ref{Dq_All} (b) behave very similarly. In both 
cases, one observes that at least the first three momenta feature a plateau 
with $D_q(t) \approx \text{const}$, which is a clear signature of diffusion and 
confirms our earlier conclusion. Note that the slight increase of $D_q(t)$ in 
Fig.\ \ref{Dq_All} (a) is not necessarily a finite-size effect 
\cite{karrasch2014_2, steinigeweg2015}.

Now, we come back to the case of smaller anisotropies. In Fig.\ \ref{Dq_All} 
(c), $D_q(t)$ is depicted for the integrable case $\Delta = 1$, $\Delta^\prime 
= 0$, while Fig.\ \ref{Dq_All} (d) shows the non-integrable case $\Delta = 
\Delta^\prime = 1$. For $\Delta^\prime = 0$, one clearly observes that the 
diffusion coefficient increases with time for all $q \geq 0$. Moreover, even 
for the smallest nonzero momentum $q/(2\pi/L) = 1$, we see deviations between 
$q \neq 0$ and $q = 0$. For $\Delta^\prime \neq 0$, $D_q(t)$ behaves 
significantly different. For $q = 0$, we have $D_q(t) \approx \text{const}$,
which is accurately reproduced at least for $q/(2\pi/L) = 1$. For larger wave  
vectors, however, we are unable to find a plateau with constant $D_q(t)$. Thus, 
compared to the case of larger anisotropies [Figs.\ \ref{Dq_All} (a) and (b)], 
the hydrodynamic regime is shifted to smaller momenta if $\Delta$, 
$\Delta^\prime$ is decreased.

Based on the data in Fig.\ \ref{Dq_All}, we conclude that the real-time dynamics 
of typical states in the XXZ chain shows diffusive behavior, not only for large 
anisotropies $\Delta = 1.5$ but also for smaller $\Delta = 1$, if integrability 
is broken due to an additional next-nearest neighbor interaction $\Delta^\prime 
> 0$. This is another main result of the present paper. Note that a similar 
result is likely to appear for even smaller anisotropies, e.g., $\Delta = 
\Delta^\prime = 0.5$. However, due to a large mean free path, we are not able 
to draw reliable conclusions in this parameter regime. More details on this 
issue are given in the appendix.

\subsection{Real-space dynamics of untypical states} \label{ResUnTyp}

\begin{figure}[tb]
\includegraphics[width=0.8\columnwidth]{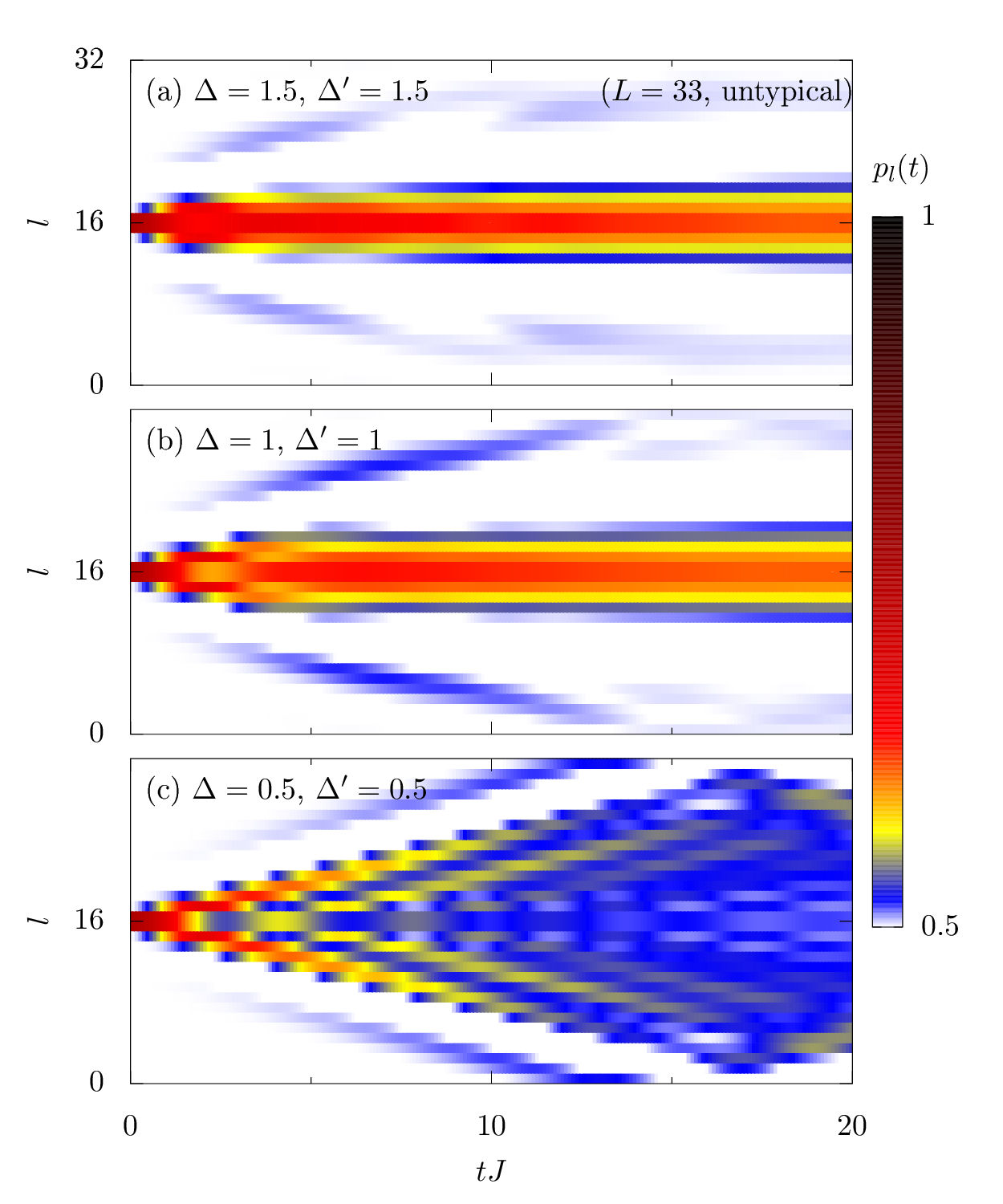}
\caption{(Color online) Time-space density plot of occupation numbers 
$p_l(t)$for an \textit{untypical} initial state $\ket{\psi(0)}$ in the XXZ 
spin-$1/2$ chain with $L = 33$ sites and different anisotropies $\Delta = 
\Delta^\prime = 1.5$, $1$, and $0.5$ [(a) - (c)].} \label{DichteC_Eq_Pic}
\end{figure}

Now, we turn to our study of \textit{untypical} initial states, where the
coefficients $c_k$ in Eq.\ \eqref{initialState} are all chosen to be  
equal. Figure \ref{DichteC_Eq_Pic} shows a time-space density plot of 
occupation numbers $p_l(t)$ for a chain with $L = 33$ sites. Completely 
analogous to Fig.\ \ref{DichteC_Ty_Pic}, panels (a) - (c) show results for 
$\Delta = \Delta^\prime = 1.5$, $1$, and $0.5$. First of all, one observes that 
the time dependence of the density profiles strongly differs from the case of 
typical initial states. On the one hand, for large interactions $\Delta$ [see 
Fig.\ \ref{DichteC_Eq_Pic} (a)], the broadening is basically frozen and the 
density profile is very narrow even at times $tJ = 20$, similar to 
\cite{gobert2005}. On the other hand, for small interactions $\Delta$ [see 
Fig.\ \ref{DichteC_Eq_Pic} (c)], one observes pronounced jets which propagate 
freely until they eventually hit the boundary at times $tJ \sim 20$. Such a 
behavior of untypical states has been already found for the integrable model 
$\Delta^\prime = 0$ \cite{steinigeweg2017_1}. Our present results clearly show 
that this behavior is stable against perturbations $\Delta^\prime \neq 0$.

\section{Properties of Typical and Untypical States} \label{properties}

In the following, we intend to shed light onto the properties of typical and 
untypical initial states, in order to provide possible explanations for the 
large differences in the real-time dynamics. As a starting point, we first 
analyze the states with respect to their local density of states. 

\subsection{Local density of states}

The local density of states (LDOS) $P(E)$ of a state $\ket{\psi}$, as well as 
the density of states (DOS) $\Omega(E)$ of a Hamiltonian $\mathcal{H}$, is 
given by
\begin{align}
P(E) &= \sum_n |\braket{n|\psi}|^2 \, \delta(E-E_n)\ , \\
\Omega(E) &= \sum_n \delta(E-E_n)\ ,
\end{align}
where $\ket{n}$ are the eigenvectors of $\mathcal{H}$ with corresponding 
eigenvalues $E_n$. While $P(E)$ and $\Omega(E)$ can be calculated using ED of 
small systems, we proceed differently here and employ a numerical 
approach \cite{hams2000, jin2016}. This approach relies again on the real-time 
propagation of pure state and, for $\Omega(E)$, on the concept of typicality. 
Details on the numerical calculation of $P(E)$ and $\Omega(E)$ can be found in 
the appendix.

In Fig.\ \ref{DOS_N20}, the DOS of $\mathcal{H}$ with $L = 24$ sites is shown 
for both, an integrable ($\Delta = 1.5$, $\Delta^\prime = 0$) and a 
non-integrable ($\Delta = \Delta^\prime = 1.5$) case. Note that $L=24$ is 
sufficient to capture the overall shape of the DOS. In both cases, $\Omega(E)$ 
has a broad Gaussian-like shape \cite{jin2016}. In addition, the LDOS $P(E)$ 
is shown for a typical state with random coefficients $c_k$ and an untypical 
state where all $c_k$ are the same. For the typical state, $P(E)$ apparently 
coincides with the DOS of $\mathcal{H}$. This fact also reflects that a typical 
state imitates the high-temperature statistical ensemble. In contrast, for an 
untypical state, $P(E)$ is sharply peaked at the upper border of the spectrum. 
This fact clearly shows that an untypical state does not imitate the 
high-temperature statistical ensemble. Moreover, since in the gapped phase 
$\Delta > 1$ the dynamics at the spectral border is expected to be insulating, 
this fact provides a reasonable explanation for the frozen density profiles in 
Fig.\ \ref{DichteC_Eq_Pic} (a).

\begin{figure}[tb]
\centering
\includegraphics[width=\columnwidth]{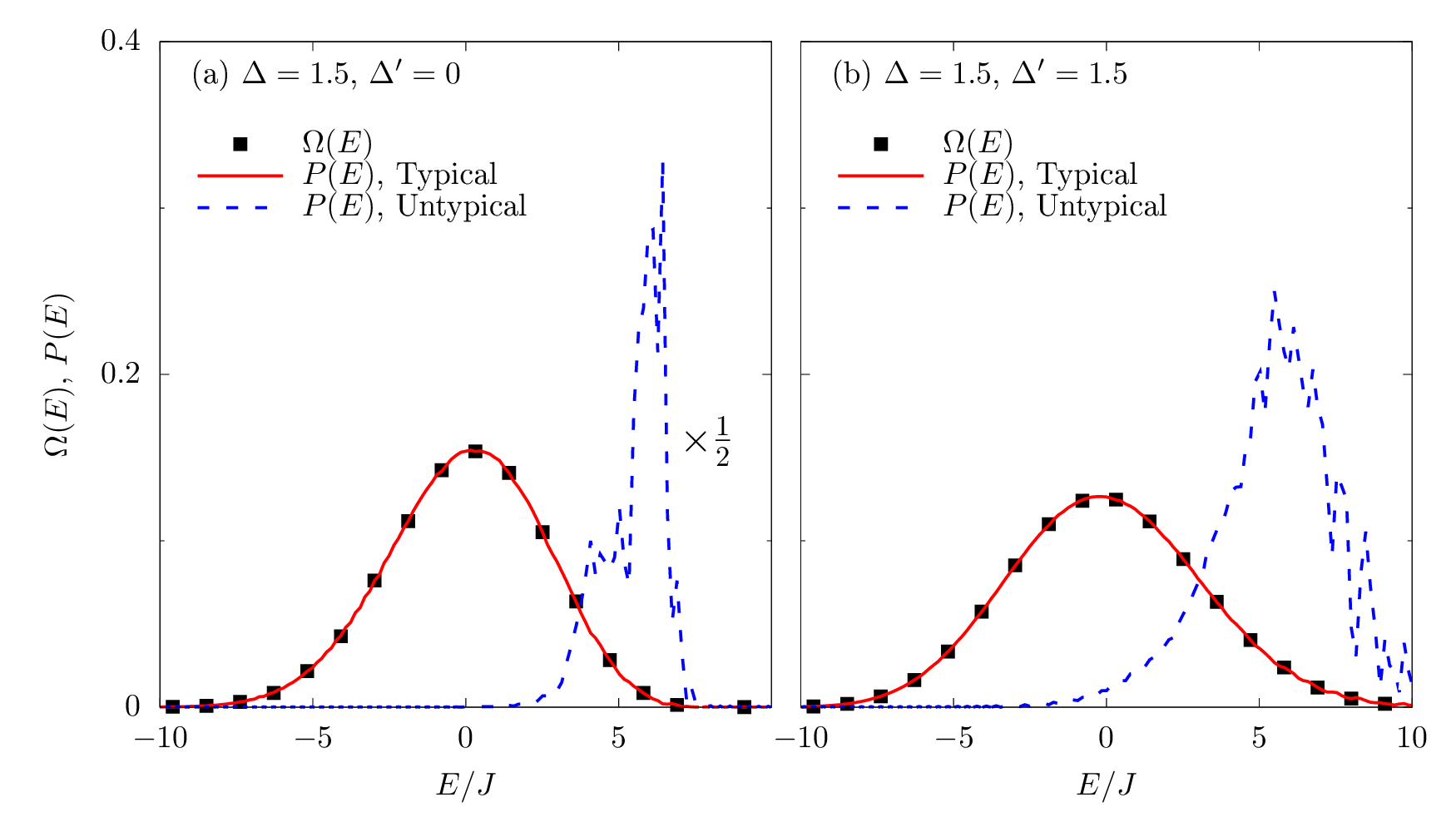}
\caption{(Color online) Density of states $\Omega(E)$ of $\mathcal{H}$ with
$L = 24$ sites and local density of states $P(E)$ for states with random $c_k$  
and equal $c_k$. Other parameters: (a) $\Delta = 1.5$, $\Delta^\prime = 0$; (b) 
$\Delta = \Delta^\prime = 1.5$.} \label{DOS_N20}
\end{figure}

\subsection{Internal randomness and entanglement}

So far, we have only distinguished between typical states, which are completely 
random, and untypical states, where the coefficients $c_k$ are all equal. At 
this point, we also analyze the role of the amount of \textit{internal 
randomness}. Moreover, we are interested in the influence of this randomness on 
the \textit{entanglement} of our non-equilibrium states. 

In order to measure the entanglement entropy \cite{schollwoeck2005} (EE) of a
given state $\ket{\psi}$, we divide our system into a left part $A$ and a right 
part $B$ of equal size. Accordingly, we write $\ket{\psi}$ as
\begin{equation}
\ket{\psi} = \sum_{i=1}^{d_A} \sum_{j=1}^{d_B}
\psi_{i,j} \ket{i} \otimes \ket{j}\ , 
\end{equation}
where $d_{A}$, $d_{B}$ are the Hilbert-space dimensions of $A$, $B$ and 
$\lbrace \ket{i} \rbrace$, $\lbrace \ket{j} \rbrace$ are orthonormal product 
bases of $A$, $B$. The reduced density matrix $\rho_A$ of part A is then given 
by 
\begin{equation}
\rho_A = \text{Tr}_B \ket{\psi}\bra{\psi}\ , 
\end{equation}
where the states $\ket{j}$ from part $B$ are traced out, 
\begin{equation}
\braket{i|\rho_A|i^\prime} = \sum_{j = 1}^{d_B} \psi_{i,j}
\, \psi_{i^\prime,j}^\ast\ .
\end{equation}
By construction, the reduced density matrix $\rho_A$ has $d_A$ eigenvalues 
$\omega_\alpha$ with $\sum_\alpha \omega_\alpha = 1$. These eigenvalues 
are then used to compute the EE, which is defined as
\begin{equation}
S = - \text{Tr} [\rho_A \log_2 \rho_A ] = -\sum_{\alpha=1}^{
d_A} \omega_\alpha \log_2 \omega_\alpha\ . 
\end{equation}

\begin{figure}[tb]
\centering
\includegraphics[width = 0.9\columnwidth]{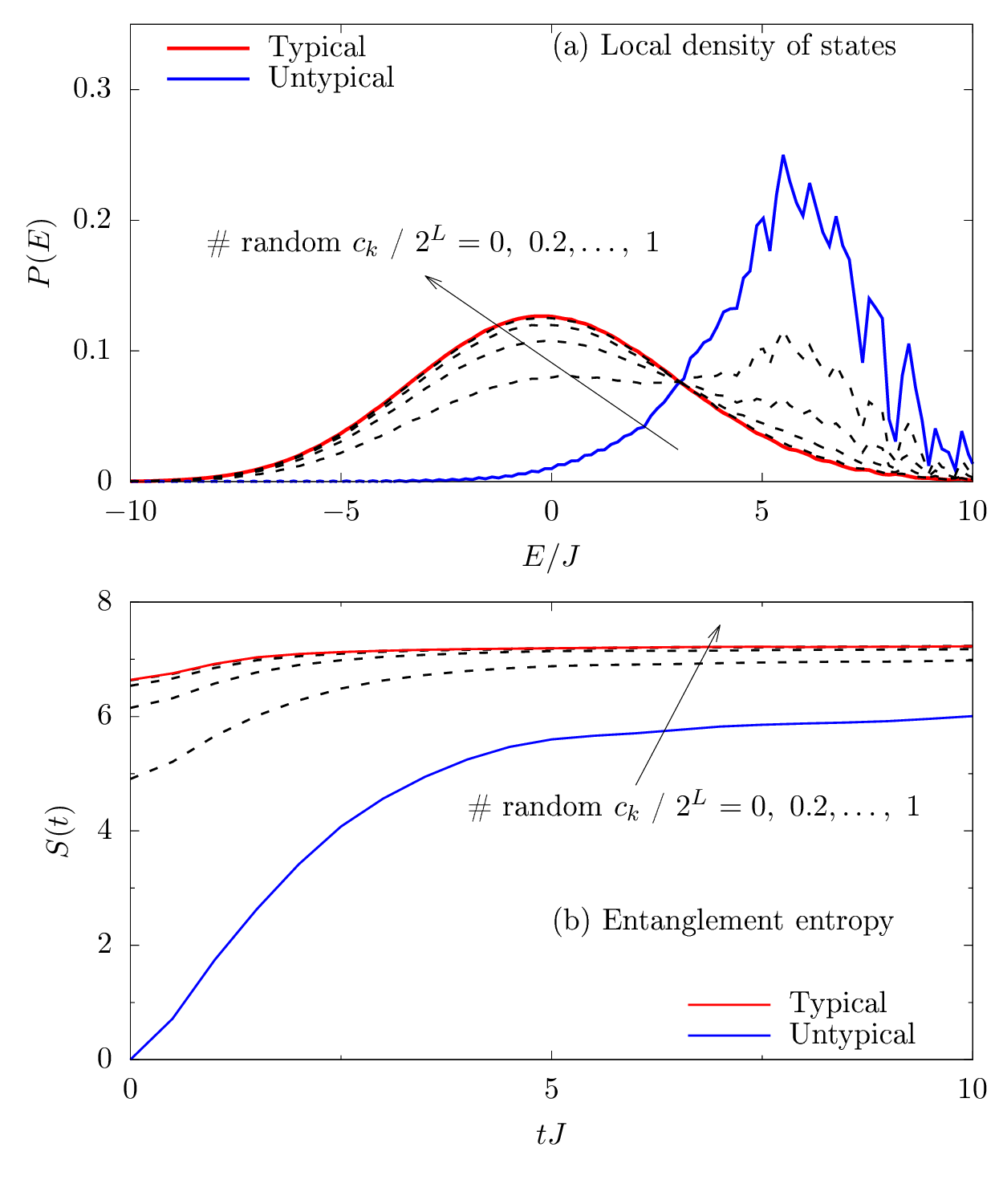}
\caption{(Color online) (a) Local density of states $P(E)$ for states with a
different number of random coefficients $c_k$ for a chain with $L = 24$ sites.
(b) Corresponding entanglement entropy $S(t)$ for a system with $L = 16$ sites. 
In both cases, we have $\Delta = \Delta^\prime = 1.5$.} \label{EE_Pic}
\end{figure}

Before we discuss the EE below, the LDOS $P(E)$ is depicted in Figs.\ 
\ref{EE_Pic} (a) and (b) for states where the percentage of random coefficients 
$c_k$ is varied between $0\%$ and $100\%$. As before, $L=24$ sites are 
sufficient. First, one observes that $P(E)$ becomes continuously broader for 
increasing randomness. In fact, for approximately $60\%$ random coefficients, 
$P(E)$ already has a pronounced Gaussian shape and is almost identical to the 
LDOS of a completely random state or the DOS $\Omega(E)$ of the Hamiltonian.

In Fig.\ \ref{EE_Pic} (b) the corresponding EE is now depicted for $L = 16$ 
sites, as obtained from ED. One observes that $S(t)$ monotonically increases at 
short times, until it eventually turns into a plateau with $S(t) \approx 
\text{const}$. Moreover, this saturation value increases with the number of 
random coefficients $c_k$ \cite{Page1993}. For an untypical state, where 
all $c_k$ are the same, we see that $S(0) = 0$, which confirms that 
$\ket{\psi}$ can be written as a product state [cf.\ Eq.\ \eqref{ProdState}].

Comparing Figs.\ \ref{EE_Pic} (a) and (b), it is evident that for our 
non-equilibrium states either a broad LDOS and high EE or a narrow LDOS and low 
EE occur simultaneously. Thus, low EE could be another explanation for the 
dynamics observed in in Fig.\ \ref{DichteC_Eq_Pic}. This possibility is 
examined below. Note that the anisotropies in Fig.\ \ref{EE_Pic} have been set 
to $\Delta = \Delta^\prime = 1.5$. We have checked, however, that the 
qualitative behavior of $P(E)$ and $S(t)$ is independent of the specific choice 
of $\Delta$ and $\Delta^\prime$ and system size $L$.

\subsection{Random product state}

As a final test to what extend internal randomness, entanglement, and LDOS
influence the real-time dynamics of our initial states, we now define a
convenient state
\begin{equation}\label{TPS}
\ket{\psi_P} = \sum_{ij} c_{ij} \ket{i} \otimes \ket{\uparrow} \otimes
\ket{j}\ ,
\end{equation}
where $c_{ij} = c_i c_j$ are complex coefficients and the sum runs over all 
states $\ket{i}$ and $\ket{j}$ of the left and right half of the chain 
respectively. By construction, $\ket{\psi_P}$ is a \textit{product state} and 
the initial density profile is identical to the class of states defined in Eq.\ 
\eqref{initialState}. Concerning the internal randomness, however, the
construction of $\ket{\psi_P}$ only involves $\sim 2^{L/2}$ random numbers, 
which is considerably less compared to a typical state with $2^L$ independent 
random coefficients.

\begin{figure}[tb]
\centering
\includegraphics[width = 0.8\columnwidth]{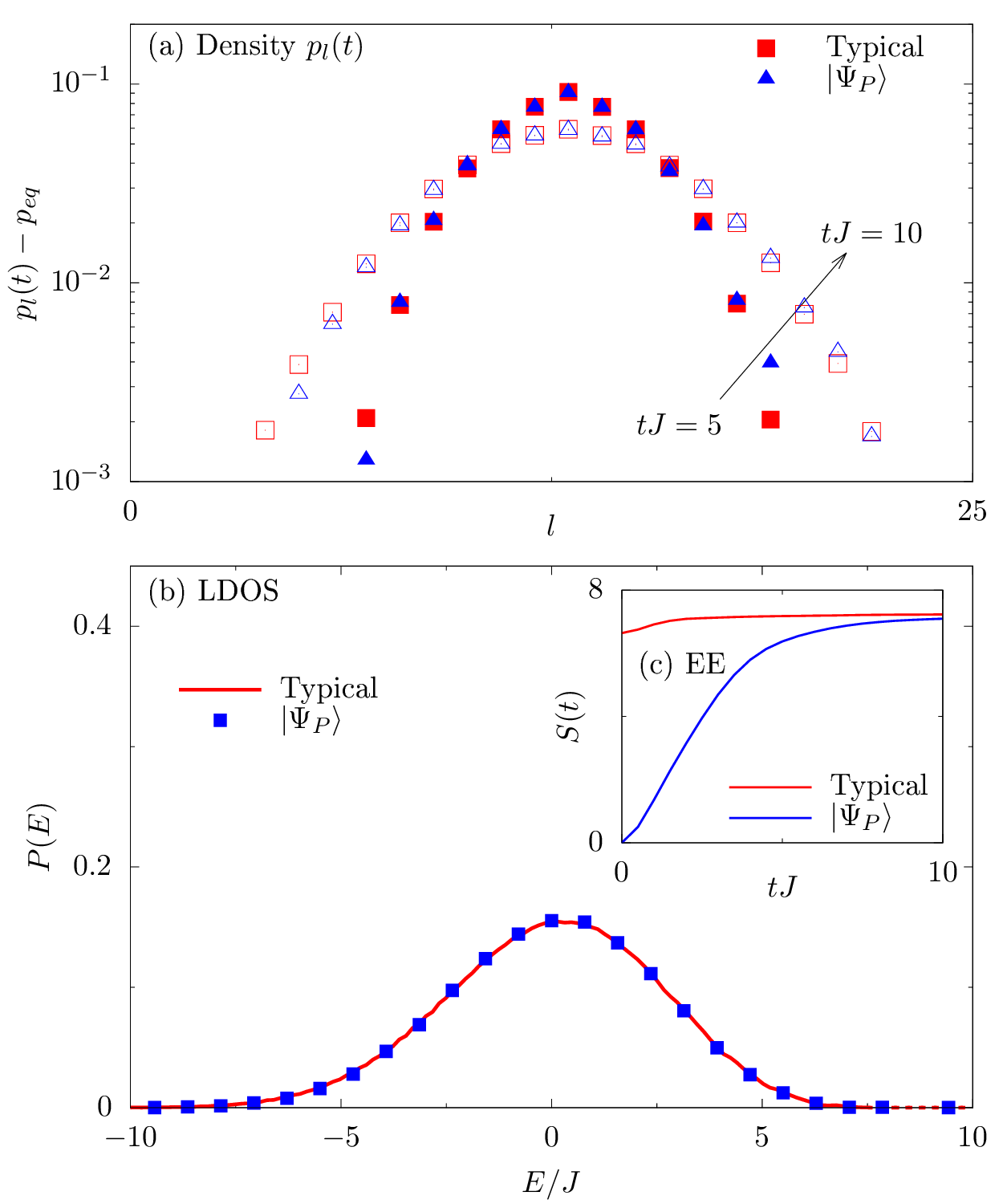}
\caption{(Color online) (a) Comparison of density profiles $p_l(t)$ for a 
typical state [see Eq.\ \eqref{initialState}] and a product state 
$\ket{\psi_P}$ [see Eq.\ \eqref{TPS}] at fixed times $t J = 5$ and $10$. (b) 
Local density of states $P(E)$ of these states for a chain with $L = 24$ sites. 
(c) Corresponding entanglement entropy $S(t)$ for a system with $L = 16$ sites. 
In all cases, we have $\Delta = 1.5$ and $\Delta^\prime = 0$.} \label{TPS_Pic}
\end{figure}

In Fig.\ \ref{TPS_Pic} (a), the density profiles $p_l(t)$ of a typical state and 
a state $\ket{\psi_P}$ according to Eq.\ \eqref{TPS} are depicted for fixed 
times $tJ = 5$ and $10$. We restrict ourselves to the integrable case with 
$\Delta = 1.5$, $\Delta^\prime = 0$ and $L = 26$ sites. In order to minimize
the dependence on the specific random initialization, the data for $\ket{ 
\psi_P}$ is averaged over $N = 20$ different initial states. Note, however,
that the total amount of random coefficients, $\sim 20 \cdot 2^{L/2}$, is still
much smaller than $2^L$. The semi-log plot in Fig.\ \ref{TPS_Pic}(a) 
illustrates that the differences between a typical state and $\ket{\psi_P}$ are 
hardly visible for all times shown here.

In Fig.\ \ref{TPS_Pic} (b), the corresponding LDOS of both states is shown. One
observes that $\ket{\psi_P}$ has a broad spectral distribution with a Gaussian 
shape which very close to the LDOS of the typical state. In Fig.\ \ref{TPS_Pic} 
(c), we also  show the entanglement entropy of both states. At $t=0$, $S(t)$ 
vanishes for $\ket{\psi_P}$ by construction. However, at longer times, $S(t)$ 
saturates at the same value as the typical state. These results suggest that 
the lack of initial entanglement is not the origin of the untypical dynamics 
observed in Fig.\ \ref{DichteC_Eq_Pic}.

\section{Conclusion}\label{CONCLsec}

To summarize, we have investigated the real-time broadening of non-equilibrium 
density profiles and, in particular, the role of the specific initial-state 
realization in non-integrable systems. To this end, we have focused on a 
particular class of initial states. This class consists of pure states and 
features initial density profiles with a pronounced peak on top of a homogeneous 
many-particle background at any temperature. As a first step, however, we have 
concentrated on the limit of high temperatures. Since this particular class of 
initial states allows for changing internal degrees of freedom without 
modifying the initial density profile, a central question has been whether and 
in how far such internal details influence the real-time and real-space 
dynamics. In this context, typicality of pure states is 
an useful concept and implies for internal randomness a dynamical behavior in 
agreement with the equilibrium correlation function. Still, this concept does  
not predict the type of transport as such and cannot be applied to initial 
states without any randomness. In particular, it cannot answer whether and for 
which initial conditions diffusion occurs in isolated systems.

As an example of a non-integrable system, we have studied the XXZ spin-$1/2$
chain, where integrability is broken due to a next-nearest neighbor interaction.
Using large-scale numerical simulations, we have first unveiled that random
initial states yield diffusive broadening in the regime of strong interactions. 
Quite remarkably, in this regime, we have found that signatures of diffusion 
are equally pronounced for the non-integrable and integrable model. Our 
numerical simulations in real space, as well as a Fourier analysis, have 
further shown the existence of diffusion for weaker interactions, as long as 
integrability is broken.

Finally, since we have observed that non-random states can lead to entirely 
different behavior, we have characterized typical and untypical states in terms 
of the amount of internal randomness, the local density of states, and the 
entanglement entropy. Here, our numerical results have suggested that different 
initial conditions lead to the same dynamical behavior if their local density 
of states is similar. The initial entanglement entropy, on the other hand, does 
not seem to be a crucial property. The latter we have demonstrated for a random 
product state.

Promising future research directions include the study of real-time dynamics of 
typical and untypical states in a wider class of non-integrable systems, e.g., 
in extended Hubbard models or spin models with disorder, also at lower 
temperatures. In addition to transport of spin and charge, it would also be
interesting to investigate energy dynamics as well.   

\section*{Acknowledgments}
We thank the DFG Research Unit FOR 2692 (Bielefeld, J\"ulich, Oldenburg, 
Osnabr\"uck) for very fruitful discussions. Additionally, we gratefully 
acknowledge the computing time, granted by the ``JARA-HPC Vergabegremium'' and 
provided on the ``JARA-HPC Partition'' part of the supercomputer ``JUQUEEN'' 
\cite{stephan2015} at Forschungszentrum J\"ulich.

\appendix

\section{Influence of initial peak height} \label{Secag0}

In the main text, we have focused on initial states $\ket{\psi(0)}$ with the 
maximum amplitude $p_{L/2}(0) = 1$ possible. For completeness, let us also 
discuss here whether or not the non-equilibrium dynamics depends on this 
particular choice. By choosing $a>0$ in the definition of our initial states 
[see Eq.\ \eqref{initialState}], it is possible to construct states with 
$p_{L/2}(0) < 1$, which are in this sense closer to equilibrium. Note that 
$p_{l\neq L/2} (0) = p_{\text{eq}} = 1/2$ is unaffected by $a > 0$.

First, it is instructive to show how the size of the initial peak $p_{L/2}(0)$ 
in the middle of the chain is controlled by the parameter $a$. To this end, the 
following calculation can be performed.
\begin{align}
\frac{\bra{\psi} n_{L/2} \ket{\psi}}{\braket{\psi|\psi}}
&= \frac{\bra{(n_{L/2}-a) \Phi} n_{L/2} \ket{(n_{L/2}-a) \Phi}}{\braket{ 
(n_{L/2}-a) \Phi|(n_{L/2}-a) \Phi}} \\[0.1cm]
&= \frac{\bra{\Phi} (n_{L/2} -a) n_{L/2} (n_{L/2}-a) \ket{\Phi}}{\bra{\Phi} 
(n_{L/2} -a)(n_{L/2}-a) \ket{\Phi}} \\[0.1cm]
&= \frac{(1-a)^2 \bra{\Phi} n_{L/2} \ket{\Phi}}{(1-2a) \bra{\Phi} n_{L/2}  
\ket{\Phi} + a^2 \braket{\Phi|\Phi}}
\end{align}
In the last step, we have multiplied out brackets and used the projection
property $n_{L/2}^3 = n_{L/2}^2 = n_{L/2}$. Since $\bra{\Phi} n_{L/2} \ket{\Phi}
= \braket{\Phi|\Phi}/2$, one therefore finds that $p_{L/2}(0)$ does not depend
linearly on $a$ but rather follows
\begin{align}
p_{L/2}(0) = \frac{\bra{\psi} n_{L/2} \ket{\psi}}{\braket{\psi| \psi}} = 
\frac{(1-a)^2}{(1-a)^2 + a^2}\ . 
\end{align}
It follows that for $a = 0$ we have $p_{L/2}(0) = 1$, whereas for $a = 0.5$ we 
have $p_{L/2}(0) = p_{\text{eq}} = 0.5$. 

\begin{figure}[tb]
\centering
\includegraphics[width=\columnwidth]{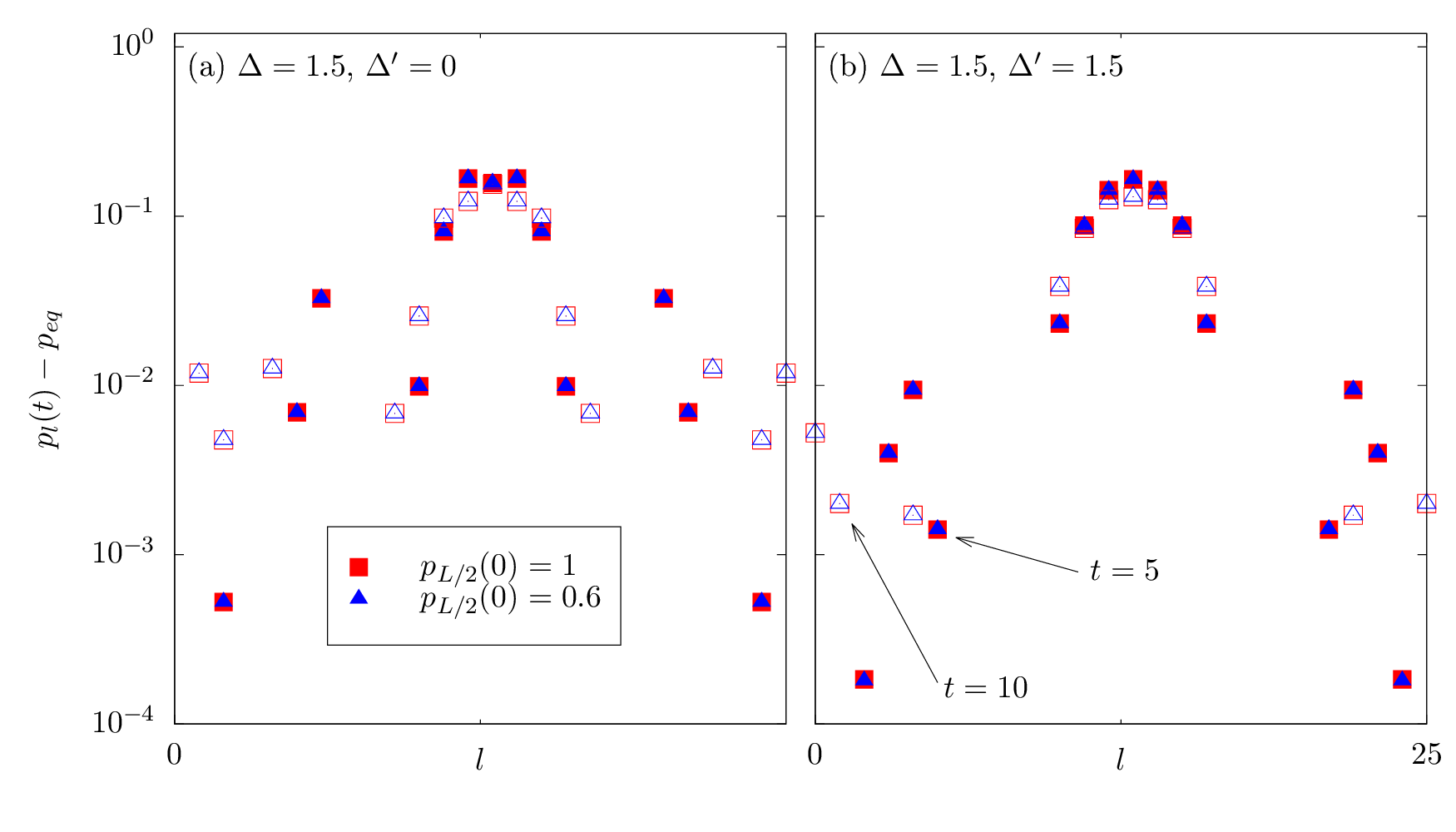}
\caption{(Color online) Density profile $p_l(t)$ for untypical initial states 
with $p_{L/2}(0) = 1$ and $p_{L/2}(0) = 0.6$ at fixed times $tJ = 5$ (filled 
symbols), $tJ = 10$ (open symbols) for a chain with $L = 26$ sites and 
different anisotropies: (a) $\Delta = 1.5$, $\Delta^\prime = 0$; (b) $\Delta = 
1.5$, $\Delta^\prime = 1.5$. For a meaningful comparison, the data for 
$p_{L/2}(0) = 0.6$ is multiplied by a factor $5$.} 
\label{Dichte_Times_peaks_pic}
\end{figure}

We now present the simulation results. Here, we focus on the case of untypical 
states, i.e., all $c_k$ are equal, and compare the dynamics of a state with 
$p_{L/2}(0) = 1$ and a state with $p_{L/2}(0) = 0.6$. In Fig.\ 
\ref{Dichte_Times_peaks_pic}, the resulting density profiles $p_l(t)$ are shown 
for $L = 26$ and different anisotropies $\Delta$, $\Delta'$, at fixed times $tJ
= 5$, $10$. For a meaningful comparison, the data for $p_{L/2}(0) = 0.6$ is 
multiplied \cite{remark} by an overall scaling factor $5$. Remarkably, after 
this simple renormalization, the data for $p_{L/2}(0) = 1$ and $p_{L/2}(0) = 
0.6$ exactly coincide with each other.

This illustrates that for an untypical state the dynamics of $p_l(t)$ is
independent of the specific initial value $p_{L/2}(0)$. In particular, by  
changing the parameter $a > 0$, it is not possible to change the dynamical 
behavior of untypical states depicted in Fig.\ \ref{DichteC_Eq_Pic} in the main 
text of this paper. Although not shown here explicitly, we have found that this 
independence of the parameter $a$ applies to typical states as well. Note that 
this independence can be also understood analytically for so-called 
binary operators \cite{richter2017}.

\begin{figure}[b]
\centering
\includegraphics[width=\columnwidth]{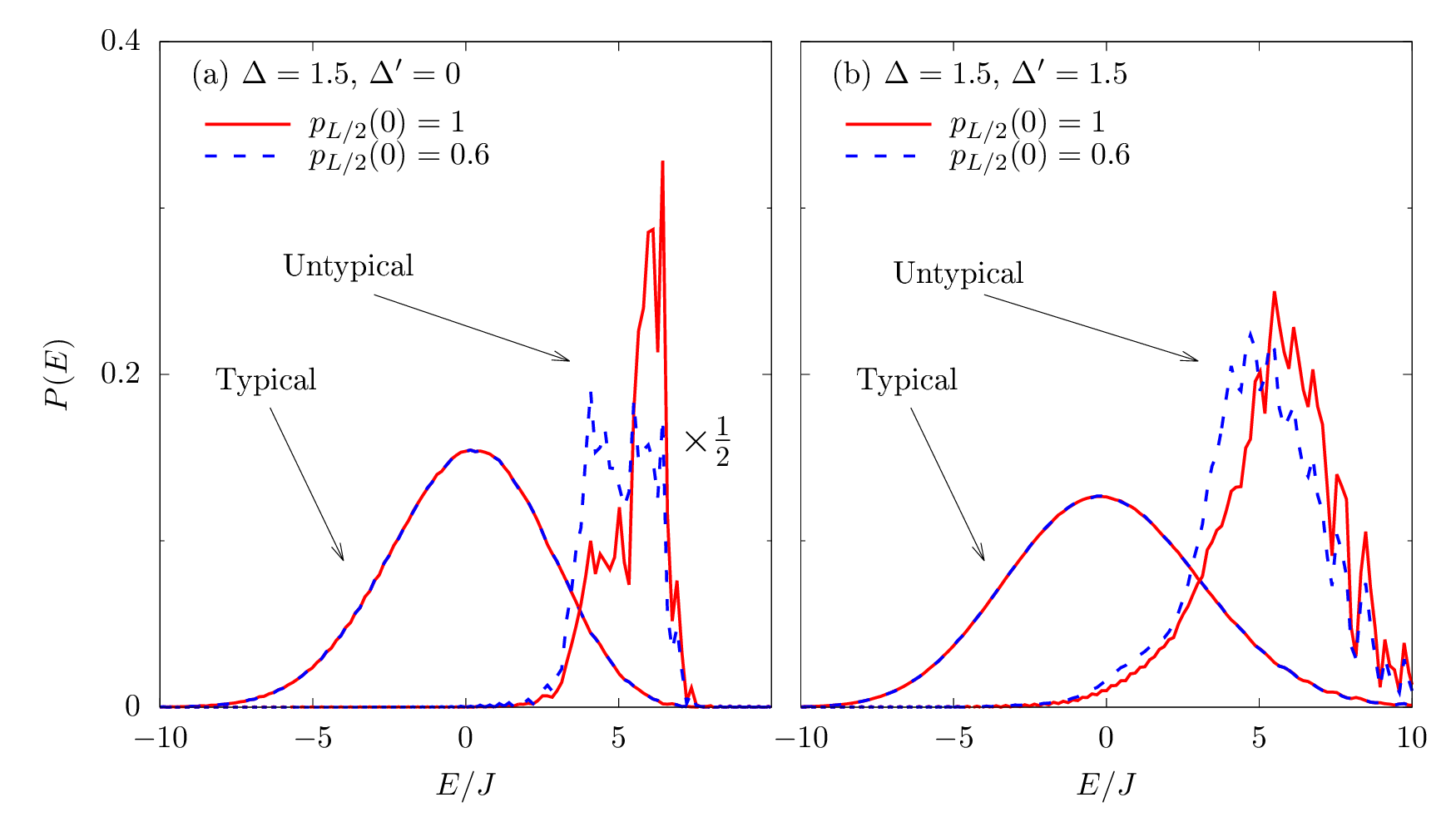}
\caption{(Color online) Local density of states $P(E)$ for typical and untypical
states with $p_{L/2}(0) = 1$ (solid lines) and $p_{L/2}(0) = 0.6$ (dashed 
lines). We use $L = 24$ in both cases. Other parameters: (a) $\Delta = 1.5$, 
$\Delta^\prime = 0$; (b) $\Delta = \Delta^\prime = 1.5$.
}
\label{DOS_N20_peak06}
\end{figure}

Finally, let us comment on the influence of $a>0$ on the LDOS $P(E)$. In Fig.\
\ref{DOS_N20_peak06}, we show $P(E)$ for typical as well as untypical initial  
states and compare the case of maximum amplitude $p_{L/2}(0) = 1$ to the case 
of $p_{L/2}(0) = 0.6$. One observes that, although the spectral weight is 
slightly redistributed compared to the case of $a = 0$, $P(E)$ is almost 
unaffected by a nonzero parameter $a > 0$. Thus, irrespective of the initial 
amplitude $p_{L/2}(0)$, a typical state has a broad Gaussian LDOS, whereas an 
untypical state goes along with a narrow LDOS at the upper border of the 
spectrum.

\section{Dynamics for small anisotropies}

In the main part of this paper, we found that the non-equilibrium dynamics of 
of $p_l(t)$ is diffusive in the regime of strong anisotropies $\Delta$ 
and $\Delta^\prime$, irrespective of the model being integrable or 
non-integrable. Furthermore, we argued that diffusion also emerges for smaller 
anisotropies, as long as integrability is broken, i.e.\ $\Delta^\prime > 0$.

\begin{figure}[tb]
\centering
\includegraphics[width=0.9\columnwidth]{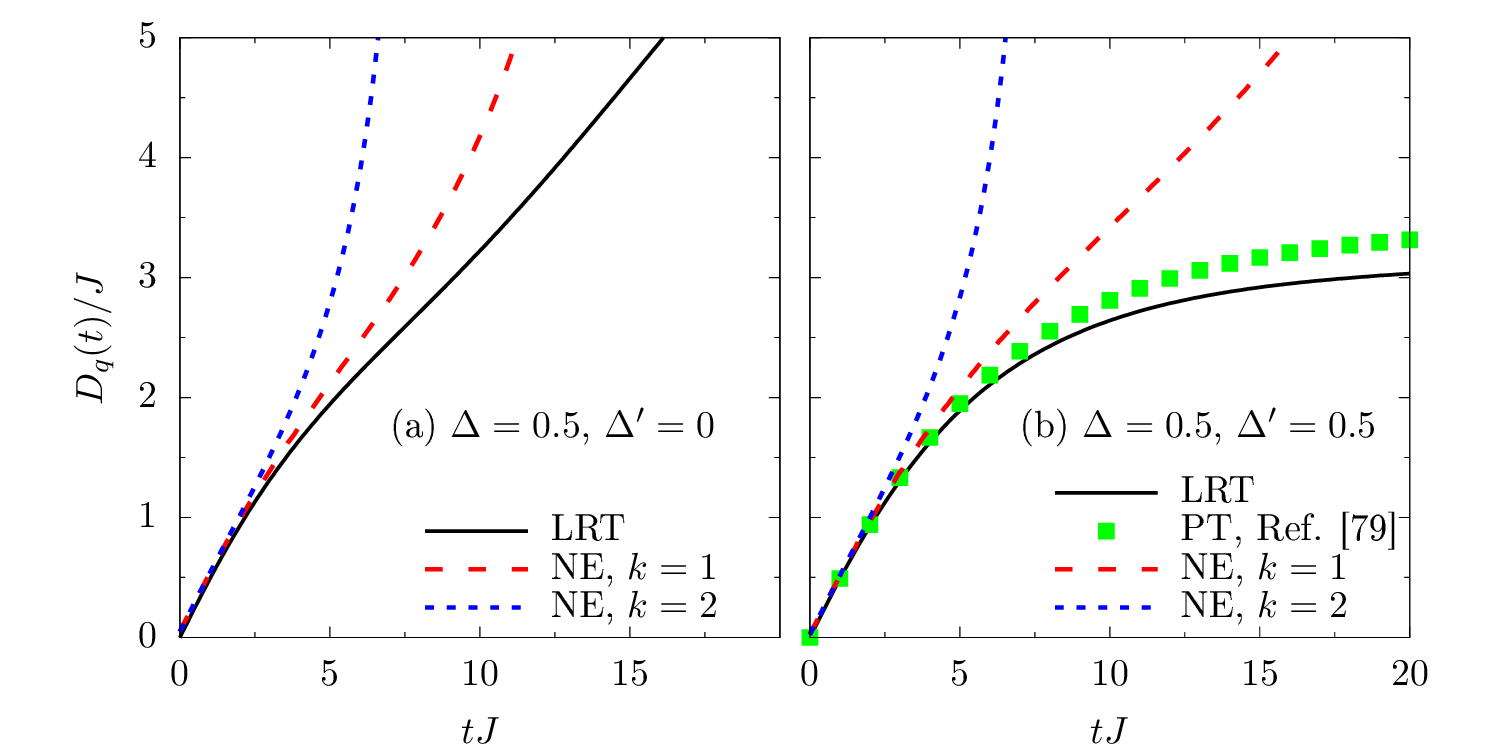}
\caption{(Color online) Generalized diffusion coefficient $D_q(t)$, obtained 
from the non-equilibrium density profiles according to Eq.\ \eqref{GenDiffC} 
for momenta $q/(2\pi/L) = k$, $L = 36$. As a comparison, $D_{q=0}(t)$ according 
to LRT is shown for $L = 33$. Moreover, we also depict data obtained by 
perturbation theory (PT) \cite{steinigeweg2011_2}. Other parameters: (a) 
$\Delta = 0.5$, $\Delta^\prime = 0$. (b) $\Delta = 0.5$, $\Delta^\prime = 0.5$.}
\label{Fig14}
\end{figure}

Let us briefly comment on the regime of small interactions 
$\Delta, \Delta^\prime < 1$. In Fig.\ \ref{Fig14}, the generalized diffusion 
coefficient $D_q(t)$, as obtained from the non-equilibrium dynamics, is shown 
for momenta $k = q/(2\pi/L) = 1,\ 2$ and anisotropies $\Delta = 0.5$, 
$\Delta^\prime = 0$ as well as $\Delta = 0.5$, $\Delta^\prime = 0.5$. For 
comparison, we also depict the diffusion coefficient $D_{q=0}(t)$, i.e. 
calculated from LRT. Concerning the non-integrable model in Fig.\ \ref{Fig14} 
(b) we observe that for $q = 0$, $D(t)$ eventually reaches a constant plateau 
at times $tJ \sim 20$. However, we are unable to find such a time-independent 
regime for any $q \neq 0$. Nevertheless, we argue that these results by no means 
rule out the possibility of diffusion. In fact, it turns out that in this 
parameter regime, the mean free time as well as the corresponding mean free 
path, are too long to draw reliable conclusion. Thus, although our data 
provides no clear evidence, they strongly suggest the emergence of diffusion in 
the thermodynamic limit also in the regime of weak interactions, as long as 
$\Delta^\prime > 0$. This conclusion is further supported by the comparison 
with the integrable case, as shown in Fig.\ \ref{Fig14} (a). Here, transport is 
clearly ballistic, $D(t) \propto t$, and at least for $k = 0$ and $k = 1$, there 
are distinct differences between the integrable and the non-integrable model.

\section{Averaging over initial states}

We briefly discuss the accuracy of our pure-state approach. For a
typical initial state, the real and imaginary part of the coefficients $c_k$ 
are drawn randomly from a Gaussian distribution with zero mean. Therefore, the 
resulting dynamics naturally depends on the specific realization of these 
random numbers. In order to reduce this dependence, we may average over $N > 1$ 
different initializations.

\begin{figure}[b]
\centering
\includegraphics[width = 1\columnwidth]{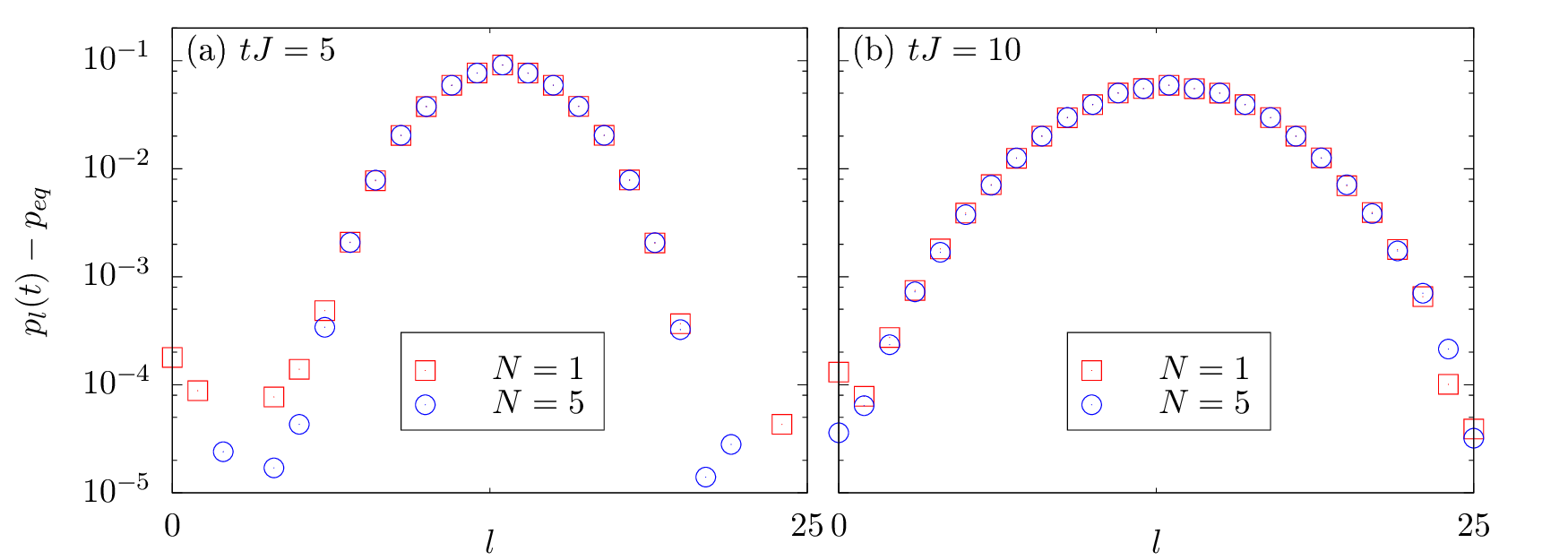}
\caption{(Color online) Density profile $p_l(t)$ with respect to site $l$ at 
fixed times $tJ = 5,\ 10$ for the integrable case $\Delta = 1.5$, 
$\Delta^\prime = 0$ and system size $L = 26$, shown in a semi-log plot. Data
for $N = 1$ random state are compared to data averaged over $N = 5$ random 
states.} \label{Fig15}
\end{figure}

In Fig.\ \ref{Fig15}, the density profile $p_l(t)$ is depicted for fixed times 
$tJ = 5$, $10$ for a chain with $L = 26$ sites and anisotropies $\Delta = 1.5$,
$\Delta^\prime = 0$. Data for $N = 1$ random state are compared to data which 
is obtained by averaging over $N = 5$ random configurations. While the Gaussian 
shape is already visible for $N = 1$, deviations from this Gaussian form at the 
boundaries are slightly reduced for the averaged data. However, these 
differences are very small and do not influence the general result. It is 
therefore sufficient to only consider $N = 1$, as done throughout the main text 
of this paper. We note that, according to typicality, errors decrease 
exponentially with increasing system size such that averaging becomes even less 
important for our large systems with $L = 36$ sites. 

\section{Forward propagation in real time}

In order to perform a forward propagation of pure states in real time, we 
employ two different methods, i.e., a fourth-order Runge-Kutta (RK4) scheme for 
\textit{medium} systems ($L \leq 26$) as well as a Trotter 
product formula for \textit{large} systems ($L > 26$). Here, we briefly 
summarize the working principle of both methods.

The time-dependent Schr\"odinger equation reads
\begin{equation}
 i \partial_t \ket{\psi(t)} = \mathcal{H} \ket{\psi(t)}\ , 
\end{equation}
where $\hbar = 1$ is set to unity. It is formally solved by
\begin{equation}\label{SolultionSE}
 \ket{\psi(t')} = U(t,t') \ket{\psi(t)}\ , 
\end{equation}
with $U(t,t') = e^{-i\mathcal{H}(t' -t)}$. While the exact evaluation of Eq.\ 
\eqref{SolultionSE} requires diagonalization of $\mathcal{H}$, we here use 
accurate approximations of the time-evolution operator $U(t,t')$.

Within the RK4 method, the Schr\"odinger equation is iteratively solved 
according to 
\begin{equation}
\ket{\psi(t+\delta t)} = \ket{\psi(t)} +  \ket{\psi_1} + \ket{\psi_2} + 
\ket{\psi_3} + \ket{\psi_4}\ ,
\end{equation}
where the $\ket{\psi_k}$ are computed as follows: $\ket{\psi_k} = (- i  
\mathcal{H})^k \delta t^k \ket{\psi(t)}/k!$. In order to ensure small numerical 
errors, we use a short time step $\delta t J = 0.01 \ll 1$ 
\cite{steinigeweg2014, steinigeweg2015, steinigeweg2016}.

Concerning the Trotter product-formula, we use a second-order approximation of 
the time-evolution operator $U(t,t+\delta t) = U(\delta t)$, given by
\begin{equation}
\widetilde{U}_2(\delta t) = e^{-i \frac{\delta t}{2} {\cal H}_k} \cdots e^{-i 
\frac{\delta t}{2} {\cal H}_1} e^{-i\frac{\delta t}{2} {\cal H}_1} \cdots 
e^{-i\frac{\delta t}{2} {\cal H}_k} \, ,
\end{equation}
where ${\cal H}={\cal H}_1 + \cdots +{\cal H}_k$. The approximation is bounded 
by
\begin{equation}
|| U(\delta t) - \widetilde{U}_2(\delta t) || \ll c_2 \, \delta t^3 \, ,
\end{equation}
where $c_2$ is a positive constant.

In practice, we use an XYZ decomposition for the Hamiltonian according to the 
$x$, $y$, and $z$ components of the spin operators, i.e., ${\cal H}={\cal H}_x 
+{\cal H}_y +{\cal H}_z$. The computational basis states are eigenstates of the 
$S^z$ operators. Thus, in this representation $e^{-i \delta t {\cal H}_z}$ is 
diagonal by construction, and it only changes the input state by altering the 
phase of each of the basis vectors. By an efficient basis rotation into the 
eigenstates of the $S^x$ or $S^y$ operators, the operators $e^{-i \delta t
{\cal H}_x}$ and $e^{-i \delta t {\cal H}_y}$ act as $e^{-i \delta t 
{\cal H}_z}$.

\section{Calculation of DOS and LDOS}

As discussed in the main part of this paper, it is possible to compute the 
(local) density of states by exact diagonalization. In this paper, however, we 
have relied on an alternative numerical approach to the DOS and LDOS 
\cite{hams2000, jin2016}. Again, we exploit the forward propagation 
of pure states in real time. The DOS can be written as
\begin{align}\label{DOS_Eq}
\Omega(E) &= \sum_n \delta(E-E_n)\ , \\ &= \frac{1}{2\pi}\int \limits_{ 
-\infty}^\infty e^{itE}\ \text{Tr} [ e^{-i\mathcal{H}t} ] \, \text{d}t\ ,
\end{align}
where we have used the definition of the $\delta$ function. According to the
principle of typicality, the trace in Eq.\ \eqref{DOS_Eq} can be evaluated by
\begin{align}\label{Typ_Eq}
\text{Tr} [ e^{-i \mathcal{H}t} ] \approx \bra{\Phi(0)} e^{-i\mathcal{H}t}
\ket{\Phi(0)} = \braket{\Phi(0)|\Phi(t)}\ ,
\end{align}
with a randomly drawn state $\ket{\Phi}$. Consequently, the DOS can 
approximately be written as  
\begin{equation}\label{DOSEq2}
\Omega(E) \approx C \int_{-\Theta}^{+\Theta} e^{itE} \braket{\Phi(0)|\Phi(t)}
\text{d}t\ ,
\end{equation}
with $ \braket{\Phi(0)|\Phi(-t)} = \braket{\Phi(0)|\Phi(t)}^\ast$ and some
normalization constant $C$. The energy resolution is given by $\Delta E = \pi/ 
\Theta$. Similarly, it is possible to define the LDOS $P(E)$ of a state 
$\ket{\psi}$ according to 
\begin{align}
P(E) &= \sum_n |\braket{n|\psi}|^2 \, \delta(E-E_n) \\
&= \frac{1}{2\pi} \int_{-\infty}^\infty e^{itE} \bra{\psi} e^{-it\mathcal{H}}
\ket{\psi} \text{d}t \label{PE_EQ1} \\
&\approx C \int_{-\Theta}^{+\Theta} e^{itE} \bra{\psi} e^{-it\mathcal{H}}
\ket{\psi}  \text{d}t\ \label{PE_EQ}\ .
\end{align}
Note that the concept of typicality is not needed in Eqs.\ \eqref{PE_EQ1} and 
\eqref{PE_EQ}. 

Since the above Fourier transforms of, e.g., $\braket{\psi| \psi(t)}$ formally 
require a signal from $t = -\infty$ to $t = \infty$, the approximation by Eq.\ 
\eqref{PE_EQ} with finite times $\Theta < \infty$ might lead to certain 
complications. This is in particular the case if the spectral representation of 
$\ket{\psi}$ is very sparse, i.e., if many coefficients $|\braket{n|\psi}|^2$ 
are zero. Then, the function $\braket{\psi|\psi(t)}$ does not necessarily 
decay, but can rather exhibit strong, almost periodic oscillations. As a 
consequence, the finite-time Fourier transform of such a signal is usually no 
smooth function, especially in the case of a high-frequency resolution, i.e., 
in the case of large cut-off time $\Theta$.

A common approach to account at least partially for this problem is the 
convolution of $\braket{\psi|\psi(t)}$ with a suitable window function. This 
window function, e.g., a Gaussian, introduces a damping of 
$\braket{\psi|\psi(t)}$ at long times and thus leads to a well-behaved Fourier 
transform. In the present paper, however, we refrain from using any kind of 
such artificial line broadening. In cases where $\braket{\psi|\psi(t)}$ is not 
decaying on a reasonable time scale, we simply restrict ourselves to short 
cutoff times $\Theta J \approx 20$, giving rise to a coarse energy resolution 
of about $\delta E/J \approx 0.15$. The resulting Fourier transform therefore 
does not necessarily produce the exact LDOS, but rather shows the general shape 
of $P(E)$. Since our aim is only to make qualitative statements about the basic 
behavior of $P(E)$, this procedure is adequate.

\newpage


\begin{thebibliography}{99}
\bibitem{remark}
The factor 5 compensates the fact that for $p_{L/2}(0) = 1$, we have 
$\sum_{l=1}^L \delta p_l(t) = 2[p_{l/2}(0) - p_{\text{eq}}] = 1$,  whereas for 
$p_{L/2} = 0.6$, we have $\sum_{l=1}^L \delta p_l(t) = 2[p_{l/2}(0) - 
p_{\text{eq}}] = 0.2$.
\end{thebibliography}
\end{document}